\documentclass{aa}  

\usepackage{graphicx}
\usepackage{txfonts}
\usepackage{xcolor}
\usepackage{lscape}
\usepackage{booktabs} 

\begin{document} 

   \title{Galaxy groups in various evolutionary stages~\thanks{Based on observations obtained at the CAHA, OHP, NOT, and TNG observatories 
(see acknowledgements for more details).}}

   \author{K.~Parra Ramos
          \inst{1,2}
          \and
          C.~Adami
          \inst{2}
          \and
          N.~Clerc
          \inst{3}
          \and
          A.~Chu
          \inst{4,5}
          \and
          F.~Durret
          \inst{5}
          \and
          G.B.~Lima Neto
          \inst{1}
          \and
          I.~M\'arquez
          \inst{6}
          \and
          L.~Paquereau
          \inst{5}
          \and
          F.~Sarron
          \inst{3}
          \and
          G.~Soucail
          \inst{3}
          \and
          P.~Amram
          \inst{2}
          \and
          Q.~Moysan
          \inst{3}
          \and
          D.~Russeil
          \inst{2}
}

\institute{Instituto de Astronomia, Geof\'isica e Ci\^encias Atmosf\'ericas, Universidade de S\~ao Paulo, Rua do Mat\~ao 1226, S\~ao Paulo/SP, Brazil
\and
Aix-Marseille Univ., CNRS, CNES, LAM, Marseille, France
\and
Universit\'e de Toulouse, CNRS, CNES, IRAP, 14 avenue E.~Belin, F-31400 Toulouse, France
\and
Stockholm University, Albanova University Center, 106 91 Stockholm, Sweden
\and
Sorbonne Universit\'e, CNRS, UMR 7095, Institut d'Astrophysique de Paris, 98bis Bd Arago, 75014, Paris, France
\and
Instituto de Astrof\'isica de Andaluc\'ia, CSIC, Glorieta de la Astronom\'ia s/n, 18008, Granada, Spain
}

\date{Received }

  \abstract
    {The formation process of galaxy groups is not yet fully understood. In particular, that of fossil groups (FGs) is still under debate. Due
   to the relative rarity of FGs, large samples of such objects are still missing. }
   {The present paper aims to analyse the properties of groups in various evolutionary stages (FGs, ``almost'' FGs, and non-FGs), and to increase the sample of FG candidates.}
   {We have spectroscopically observed galaxies in four groups and ten candidate FGs detected in the Canada France Hawaii Telescope Legacy Survey. We searched for substructures by applying the Serna--Gerbal dendrogram method to analyse the dynamical structure of each group. By applying the FIREFLY software to the continuum and PIPE\_VIS to the emission lines, we derived the stellar population properties (e.g. number and age of starbursts, metallicity, fraction contributed by the burst to the luminosity) in various regions for each group. }
   {A roughly continuous variation in properties is found between a group that is still building up (XCLASS~1330), a well-formed massive group (MCG+00-27-023), a dynamically complex non-FG (NGC~4065), and a near-FG (NGC~4104).
   We also optically confirm two FGs in the Canada France Hawaii Telescope Legacy Survey, but their X-ray luminosity is still unknown. 
   }
   {We observe that the lower the mass of the substructure, the more recent the stellar population in the considered groups. We also show an apparent lack of high-mass substructures for low-metallicity systems. These  results  are  consistent  with  the  generally  adopted model of energy transfer during interactions of the galaxies with the group and cluster potential wells. Furthermore, the fossil status of a group might be related to the large-scale environment. Therefore, studying the positions of non-FGs, near-FGs, and FGs within the cosmic web can provide insights into the process of how fossil systems come into being in the Universe.}


   \maketitle
%

\section{Introduction} 

The formation and evolution of galaxy groups and clusters remain a subject to be explored due to many open questions, with new ones arising from recent  observations made, for example, by the James Webb Space Telescope (JWST). On the one hand, many studies focus on the high-redshift Universe -- for instance, the discovery of over-massive galaxies \citep{2023Natur.616..266L,2023NatAs...7..731B} that ignited a debate about the initial mass function (IMF) in the high-redshift Universe \citep[e.g.][]{2024MNRAS.534..523C} -- or on the study of protoclusters and the properties of their galaxy members, to understand how the progenitors of current massive clusters have  evolved into the structures we observe today \citep[e.g.][]{2023ApJ...947L..24M,2024A&A...688A.146A}. On the other hand, other works focus on low and intermediate redshifts to study the evolution of the galaxy properties in groups and clusters \citep[e.g.][]{2007ApJ...670..279I,2011MNRAS.411..929G}. 
Galaxies are known to be influenced by their environment and to undergo pre-processing in groups 
\citep[see e.g.][and references therein]{Sarron+18,Kleiner+21,Bidaran+22,Lopes+24}. Their degree of evolution can be studied based on several physical properties, such as their dynamical state or star formation \citep[e.g.][and references therein]{Annunziatella+14}. 

The nature of dark matter and dark energy also remains a mystery. As a result, making predictions about the evolution of groups and clusters of galaxies beyond the present, within the framework of the $\Lambda$ cold dark matter ($\Lambda$CDM) cosmological model, still carries some uncertainties. Despite that, within the context of the hierarchical structure formation scenario,
\cite{1993Natur.363...51P} suggested the existence of an extreme class of objects called fossil groups (FGs), which can be considered as the last stage of group evolution. They were discovered by \cite{Ponman+94} and are particular groups of galaxies with high X-ray luminosities but few bright galaxies compared to typical groups or clusters of galaxies. 
\cite{Jones+03} later gave the commonly accepted definition of FGs as satisfying three conditions: they are extended X-ray sources with an X-ray luminosity of L$_{\rm X} \geq 10^{42}$ h$^{-2}_{50}$ erg~s$^{-1}$, with a brightest group galaxy (BGG hereafter) at least two magnitudes brighter than other group members ($\Delta m_{12} \geq 2.0$ mag in the R band), the distance between the two brightest galaxies being smaller than half the group virial radius. The formation of these peculiar objects and the reasons behind the low amount of optically emitting matter in them are still under debate. 
Some optical studies support the scenario that FGs are the result of high dynamical activity at high redshift, but in
an environment that is too poor for them to evolve into a cluster of galaxies through the hierarchical growth of structures. 

At X-ray wavelengths, based on Chandra X-ray observations, \cite{Bharadwaj+16} found that FGs are mostly cool-core systems, suggesting that these structures are no longer dynamically active.
However, recent observations tend to contradict the findings that FGs are dynamically relaxed systems that have not undergone recent merging events. For example, 
\cite{LimaNeto+20} detected shells around the BGG of NGC~4104 and, based on N-body simulations, showed that this FG has probably experienced a relatively recent merger between its BGG and another
bright galaxy with a mass of about 40\% of that of the BGG. More details on FGs can be found in the review by \cite{Aguerri+21}.

To make up for the lack of large samples of FGs, \cite{Adami+20} made a statistical study of FGs, extracted from the catalogue of 1371 groups and clusters detected by \cite{Sarron+18} in the Canada France Hawaii Telescope Legacy Survey (CFHTLS). The detection of these systems was based on photometric redshifts \citep{Ilbert+06}. \cite{Adami+20} found that groups with masses larger than $2.4\times 10^{14}$~M$_\odot$ had the highest probability of being FGs and discussed their location in the cosmic web relative to nodes and filaments \citep[for a similar study, see also][]{Zarattini+22}. They concluded that FGs were most probably in a poor environment, making the number of nearby galaxies insufficient to compensate for the accretion by the central group galaxy. We have obtained spectroscopy for ten of these FG candidates and confirm that the optical properties of two of them correspond to those of FGs, but X-ray data are necessary to confirm their FG nature.

Our aim in the present paper is to study the pre-processing of galaxies in groups 
by analysing their properties in various evolutionary stage structures. For this, our sample includes the still building up group XCLASS~1330, the massive group embedded in a filament MCG+00-27-023, the non-FG NGC~4065, the dynamically active FG NGC~4104 \citep{LimaNeto+20}, and the FG candidates that we found in the CFHTLS. We checked the status of each of these objects and their levels of substructure to understand their dynamical evolution. We also analysed their star formation activity through the observation of their continuum and emission lines. 

This paper is organized as follows.  
The methodology and sample selection criteria are described in Sect.~\ref{sec:methodology}. From Sect.~\ref{sec:X1330} to \ref{sec:N4104}, we present the analysis and results for each individual group of our sample. Section~\ref{sec:CFHTLS} describes the analysis of the FG candidates. Finally, the discussion and conclusions are summarized in Sect.~\ref{sec:discu}, where we describe, among other matters, the relation between different substructure properties such as age and mass. We have adopted in this paper the following cosmological parameters:  $\Omega_M = 0.286$, $\Omega_\Lambda = 0.714$, and $H_0 = 69.6\,$km~s$^{-1}$~Mpc$^{-1}$.

\section{Methodology}
\label{sec:methodology}

\subsection{Four comparison groups}
\label{sec:data}
In order to study the process of a FG formation, we selected four galaxy groups, possibly progenitors of FGs, in different evolutionary stages, from galaxy groups in an ongoing building phase (therefore potentially containing high levels of merging activity) to near-FGs, to compare their properties with those of a sample of FG candidates.

Cosmic filaments seem to host these FGs when they are not positioned too close to cosmic nodes \citep{Sarron+18}. For this reason, we selected non-FGs with a high level of substructures (SbS), located not too close to massive clusters, which are assumed to be within the cosmic nodes, and at various distances from these massive clusters. 

We used public spectroscopic data from the 18th data release of the Sloan Digital Sky Survey\footnote{https://skyserver.sdss.org/dr18} (SDSS) and the 6dF Galaxy Survey \citep[6dFGS,][]{2009MNRAS.399..683J} to obtain the redshifts of the galaxies in our galaxy group sample. To improve the completeness of the sample, we complemented the public spectroscopic data with observations from MISTRAL (Appendix~A and Table \ref{tab:NewMISTRAL}), a low-resolution single-slit spectro-imager installed on the 1.93m telescope at the Observatoire de Haute-Provence (OHP).

\subsection{Substructure detection}

We applied the Serna--Gerbal dendrogram method \citep[][hereafter SG]{Serna1996} to search for SbS within the galaxy groups.
Several SbS studies already exist in the literature for such groups (this was the basis for selecting these groups), but they are not homogeneous, since they are found by applying different methods. The use of the SG algorithm ensures that we have comparable results for each of the considered groups. SG extracts from the galaxy catalogues the dynamical SbS present in each of the considered
galaxy groups. This code was already successfully used in several works
\citep[e.g.][]{LopezGutierrez+22}. It detects SbS by computing the relative potential
energies of each pair of galaxies using the magnitude of each galaxy as a proxy for its stellar mass. It requires an input catalogue of positions, velocities, and magnitudes
(converted to masses assuming a value of the mass-to-luminosity ratio (M/L), $r$-band magnitudes being extracted in the present paper from the PanSTARRS database\footnote{https://catalogs.mast.stsci.edu/panstarrs/}),
and a defined number, $n$, of galaxies needed to build a group. Here, we most of the time chose $n = 3$. The SG test works hierarchically, seeking for lower-level SbS within a primary (more massive) detected group. 
It delivers a list of SbS labelled, for example, group~1, then groups 11 and 12, which are SbS of group 1, and so on (see Table \ref{tab:SGtable}). For each group, the SG provides an output list of galaxies belonging to each substructure, as well as the mass of each group and its crossing time.

We now explain which SbS will be discussed in the following, and why.

- Each group~1 (G1) is close to the full spectroscopic sample for a given line of sight (only a few galaxies are removed, most of the time obviously non-structure members). The G1 SbS of Table \ref{tab:SGtable} will therefore not be discussed in the following. 

- All SbS ending with the number `2' (G2, G12, G112, etc.) are 
close structures along the line of sight that do not originate from the group initial galaxy population. (1) These SbS are sometimes at very different redshifts from the main group (G2 for MCG+00-27-023, G2 and G12 for NGC~4065 in Table \ref{tab:SGtable}). These SbS dynamically unrelated to the group are therefore not discussed in the following when they are at more than 2000~km~s$^{-1}$ from the main group (typically 3 times a group velocity dispersion). (2) Other SbS are at a redshift similar to the mean group redshift and can be dynamically related to the main group galaxy population. These SbS are discussed in the following (G12, G111112 in MCG+00-27-023, G112 and G1112 in NGC~4065, and G2 and G12 in XCLASS1330 and NGC~4104).

- Now there remain only SbS ending with the number ‘1’, below the G1 level. We discuss the inner core SbS, which are the dynamically deepest embedded structures with more than three galaxies (G1111111 in MCG+00-27-023, G1111 in NGC~4065, and G11 in XCLASS1330 and NGC~4104). 
 These can be interpreted as the seeds of the four considered groups.
Other SbS ending with 1 are just intermediate level dynamical SbS and are not discussed in the following, with the exception of G111 in MCG+00-27-023 only for the dynamical analysis.

Absolute masses of individual structures are not very accurate (in particular since they depend on the value chosen for M/L), but the relative masses of the groups, normalized to the total mass of the cluster members, are reliable. The provided masses are sometimes unphysically low, simply meaning that they are too low to be computed by the code. We therefore chose to limit the masses to values larger than 10$^{12}$ M$_{\odot}$ (a typical Milky Way-class galaxy mass). The same occurs for crossing times, so we chose to limit our values to 0.1 Gyrs, the typical time resolution of the PIPES\_VIS code (see below).

\subsection{Spectral analysis}
We computed mean spectral models of the galaxies within the SG SbS when SDSS or 6dFGS spectra were available (to ensure the best possible quality, with a relatively high spectral resolution).
For possible FGs, we limited our study to only the best spectra we took (flag=4) for the BGG.
We then re-scaled all these spectra to z=0 and summed them in order to get a single spectrum per SbS,
ranging from $\sim$3800~\AA\ to $\sim$7600~\AA, representing the mean stellar population of the considered
SbS.
These spectra (see also Appendix~B) were first modelled with the FIREFLY (Wilkinson et al. 2017) spectral modelling code. Since these authors have shown that
Kroupa and Salpeter IMFs gave comparable results, we limited our analysis to a Kroupa IMF. We used the STELIB and MILES
stellar libraries, to estimate the robustness of our results. FIREFLY fits the observed spectra using multiple star formation bursts
with various luminosities, masses, metallicities, and ages. The software returns the number of bursts, and for each burst its age,
metallicity, and the fraction contributed by the burst to the luminosity and stellar mass of the galaxy. As a first step, we only fitted the continuum
since Wilkinson et al. (2017) have shown that emission lines do not strongly modify the parameters of the fit. Emission lines will be
treated later with the \texttt{PIPES\_VIS} tool.
In order to estimate physical parameters that are mainly sensitive to emission lines, we chose the \texttt{PIPES\_VIS} visualization tool
(Leung et al. 2021), based on the BAGPIPES tool (Carnall et al. 2018). 
This code allows for the
manual modelling of spectra, including a nebular emission component mainly driven by the
ionization parameter, U. The bursts were modelled assuming delayed star formation events assuming a star formation rate (SFR) decay
timescale noted as $\tau$ and a nebular component taken from Leung et al. (2021). Parameters mainly depending on the continuum were fixed to the fitted
values obtained from the FIREFLY fit (age and metallicity of the main stellar population, number of bursts, and their contribution to the total stellar mass). 

The \texttt{PIPES\_VIS} parameters mainly sensitive to emission lines are the age of the latest burst, the $\tau$ value
of the delayed SFR model, and the U excitation parameter. U allows one to change the [OIII]/H$\alpha$ ratio,
$\tau$, and the age of the latest burst act directly on the presence of emission lines. We noted a clear degeneracy
between $\tau$ and the age of the latest burst in the \texttt{PIPES\_VIS} tool. We chose to fix $\tau$ to 0.3 as this
value is in good agreement with \cite{Wilkinson2017MNRAS.472.4297W}
and allows us to model all the summed spectra that we considered. We
therefore changed the age of the latest burst in \texttt{PIPES\_VIS} to model the emission lines present in our
summed spectra.

\subsection{Galaxy spatial sampling}

The main caveat we see in our methodology is that our spectral analysis is based on spectra from the SDSS
and 6dFGS surveys, which are taken with fibers usually placed on the centre of the galaxies. It therefore excludes
part of the signal coming from the outer regions of the considered galaxies. 
Since star formation mainly
occurs in the outer regions of the galaxies, this may underestimate the emission lines really present in the
considered galaxies.
In order to estimate the amplitude of the effect, we compared the size of the galaxies in our sample (extracted
from the Pan-STARRS catalogue to have homogeneous measurements) with the size of the 6dFGS and SDSS fibers. For the
MCG+00-27-023 group, 100$\%$ of the considered galaxies are fully covered by the 6dF fiber up to a galaxy radius larger
than 3$\sigma$ of the galaxy light distribution. For the NGC~4065 and NGC~4104 groups, 72$\%$ of the considered
galaxies are fully covered by the SDSS fiber at a radius larger than 3$\sigma$ of the galaxy light distribution.
Considering the 2.7$\sigma$ radius of the galaxy light distribution, 100$\%$ of the considered
galaxies are fully covered by the SDSS fiber. For NGC~4104 itself, one of the only galaxies in our sample significantly larger than the SDSS and 6dFGS fibers, \cite{Chu+23} show that the disc of the galaxy does not exhibit emission lines.
We therefore conclude that the limited size of the fibers will only weakly affect our results.

\begin{table}[h]
\centering           
\caption{Substructures identified with the Serna--Gerbal dendrogram method in our group sample.}
\label{tab:SGtable}
\begin{tabular}{lcccc}       
\hline 
\hline 
\multicolumn{5}{l}{XCLASS 1330 group} \\
\hline
ID  &  Number &  Virial Mass    &  Cross. Time  & cz        \\
& of galaxies       &  M$_{\odot}$    &  Years        & km~s$^{-1}$    \\
\hline
G1  &   12    &$\leq$0.1 $\times$ 10$^{13}$&  $\leq$0.1 $\times$ 10$^{9}$ & 58827  \\
G2  &  ~3      &$\leq$0.1 $\times$ 10$^{13}$&  $\leq$0.1 $\times$ 10$^{9}$ & 58634  \\
G11  &    ~8   &$\leq$0.1 $\times$ 10$^{13}$& $\leq$0.1 $\times$ 10$^{9}$  & 58636  \\
G12  &    ~4   &$\leq$0.1 $\times$ 10$^{13}$& $\leq$0.1 $\times$ 10$^{9}$  & 59162  \\
\hline 
\hline 

\multicolumn{5}{l}{MCG+00-27-023 group} \\
\hline
ID  &  Number  &  Virial Mass    &  Cross. Time  & cz       \\
& of galaxies &  M$_{\odot}$    &  Years        & km~s$^{-1}$    \\
\hline 
G1  &  58   &   4.1 $\times$ 10$^{13}$  &  1.3 $\times$  10$^{9}$ & 11061  \\
G2  &   ~4   &   3.9 $\times$  10$^{13}$  &  1.8 $\times$  10$^{9}$ & 12485  \\
G11 &  52   &   2.4 $\times$  10$^{13}$  &  1.5 $\times$  10$^{9}$ & 10983  \\
G12 &  ~6    &   0.3 $\times$  10$^{13}$  &  9.2 $\times$  10$^{9}$ & 11740  \\
G111 &   42      &   2.3 $\times$  10$^{13}$  &   1.4 $\times$ 10$^{9}$ & 10945  \\               
G1111 &  27      &   1.5 $\times$ 10$^{13}$  &  1.2 $\times$  10$^{9}$ & 10880  \\
G11111  &  23      &   1.5 $\times$  10$^{13}$  &  1.1 $\times$  10$^{9}$ & 10869  \\
G111111  &          17      &   0.7 $\times$ 10$^{13}$  &  1.0 $\times$ 10$^{9}$ & 10951  \\
G111112  &        ~3       &$\leq$0.1 $\times$  10$^{13}$&  2.4 $\times$ 10$^{9}$ & 10357  \\
G1111111 &  ~8       &   0.3 $\times$ 10$^{13}$  &  1.3 $\times$ 10$^{9}$ & 11014  \\

\hline 
\hline 
\multicolumn{5}{l}{NGC~4065 group} \\
\hline
ID  &  Number  &  Virial Mass    &  Cross. Time  & cz       \\
& of galaxies &  M$_{\odot}$    &  Years        & km~s$^{-1}$    \\
\hline
G1  &   151 & 0.3 $\times$ 10$^{13}$  &  $\leq$0.1 $\times$ 10$^{9}$ & 7172   \\
G2  &   ~~7  &$\leq$0.1 $\times$ 10$^{13}$&  $\leq$0.1 $\times$ 10$^{9}$ & 13936  \\
G11  &  136   &$\leq$0.1 $\times$ 10$^{13}$& $\leq$0.1 $\times$ 10$^{9}$  & 6983   \\
G12  & ~~3   &$\leq$0.1 $\times$ 10$^{13}$& 0.15 $\times$ 10$^{9}$ & 12541  \\
G111& ~37  &$\leq$0.1 $\times$ 10$^{13}$&  $\leq$0.1 $\times$ 10$^{9}$ & 6862   \\               
G112& ~~3   &$\leq$0.1 $\times$ 10$^{13}$& $\leq$0.1 $\times$ 10$^{9}$  & 6831   \\               
G1111  & ~13      &$\leq$0.1 $\times$ 10$^{13}$& $\leq$0.1 $\times$ 10$^{9}$  & 6794   \\
G1112  &  ~~6       &$\leq$0.1 $\times$ 10$^{13}$&  $\leq$0.1 $\times$ 10$^{9}$ & 7280   \\

\hline 
\hline 
\multicolumn{5}{l}{NGC~4104 group  } \\
\hline
  
ID  &  Number  &  Virial Mass    &  Cross. Time  & cz       \\
& of galaxies &  M$_{\odot}$    &  Years        & km~s$^{-1}$    \\
\hline
G1  &    17      &  1.4 $\times$ 10$^{13}$  &  1.2 $\times$ 10$^{9}$ & 8352   \\
G2  &   ~3       &$\leq$0.1 $\times$ 10$^{13}$&  0.5 $\times$ 10$^{9}$ & 7851   \\
G11  &  ~3       &  0.7 $\times$ 10$^{13}$  &  0.5 $\times$ 10$^{9}$ & 8675   \\
G12  &    ~2       &  0.5 $\times$ 10$^{13}$  &  0.1 $\times$ 10$^{9}$ & 8525   \\
\hline 
\end{tabular}
\end{table}

\section{The XCLASS~1330 building group}
\label{sec:X1330}

The XCLASS survey is a search for X-ray emitting galaxy clusters in observations retrieved from the entire XMM-\textit{Newton} archive \citep{2021A&A...652A..12K}. Extended sources were carefully distinguished from point-like sources in order to detect cluster candidates. A thorough visual inspection was performed to eliminate sources that are not galaxy clusters and that are easily identified in shallow optical images: nearby galaxies, detector artefacts, and so on. After performing an extensive search for known redshifts in the literature, the catalogue was matched to optical photometry \citep{2019AJ....157..168D,2023MNRAS.526..323Z} 
in order to obtain photometric redshift estimates for all the $z<0.8$ remaining cases (Moysan et al. in preparation). The XCLASS sample contains more than 1600 galaxy clusters detected over a 169 deg$^2$ area spread over the entire extragalactic sky. 

\begin{figure}[ht]
\begin{center}
\includegraphics[width=0.4\textwidth]{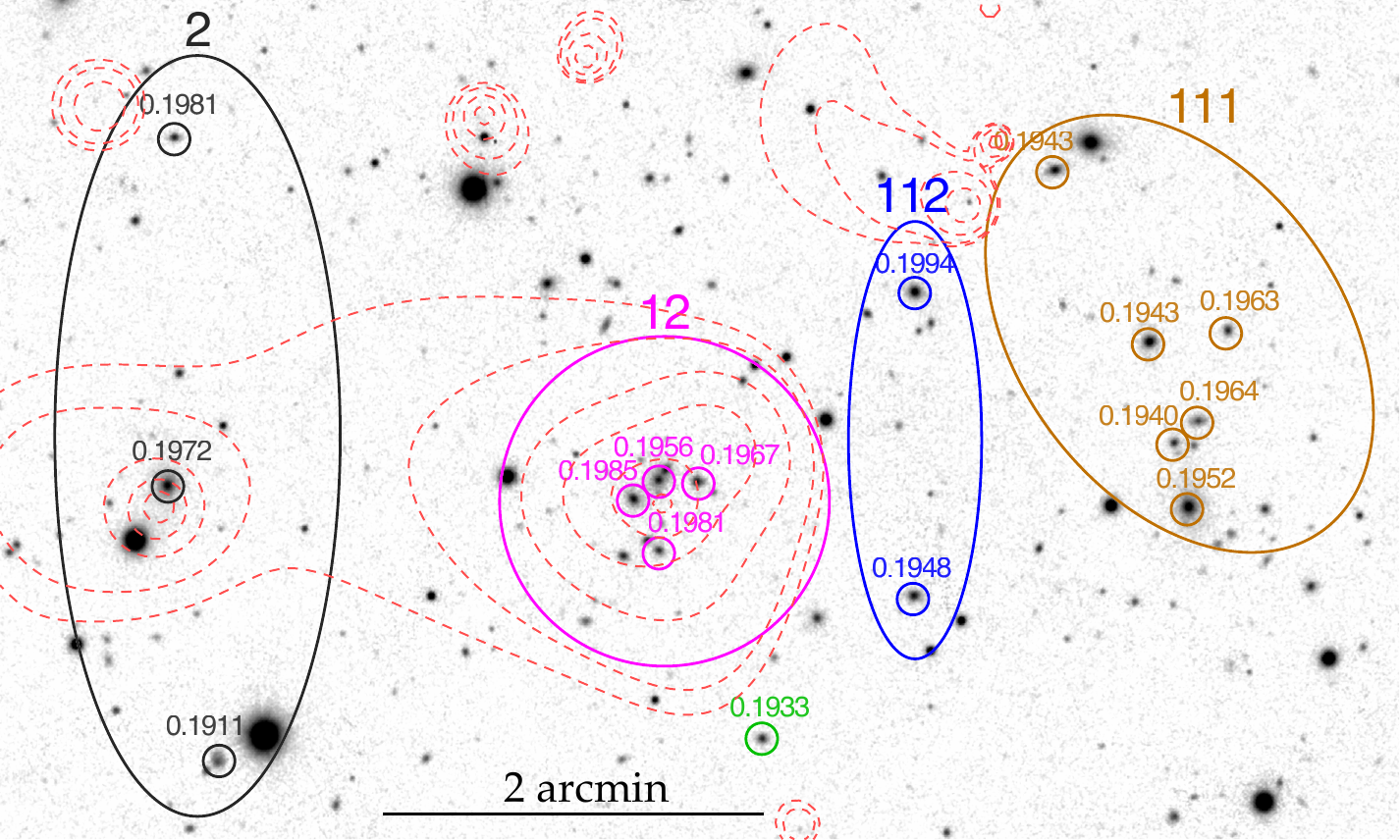}
\includegraphics[width=0.46\textwidth]{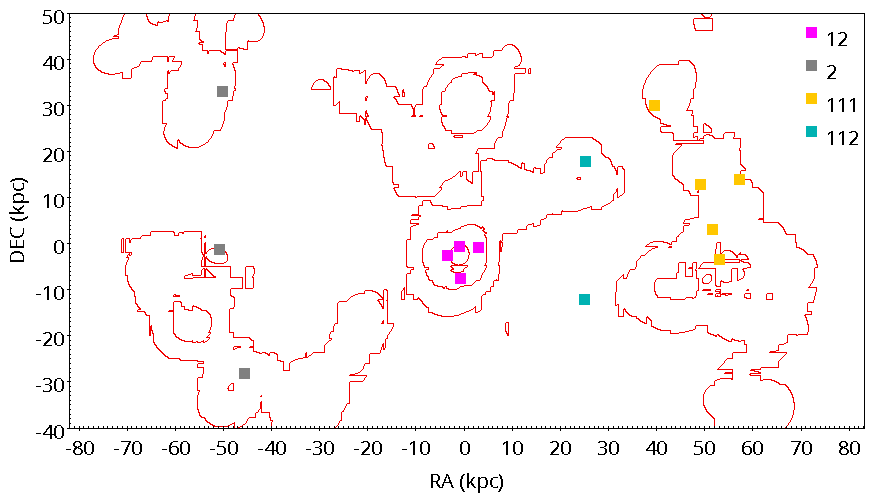}
\end{center}
\caption{Various SbS detected within the XCLASS~1330 group. \textsf{Top panel:} Substructure regions overlaid on the Pan-STARRS 
r-band image. The member galaxies with measured redshifts are also shown with the same colours as their respective region. The dashed red lines correspond to X-ray surface brightness contours (XCLASS private communication). The small green circle indicates a galaxy with a redshift within XCLASS~1330, but not associated with a particular substructure.
\textsf{Bottom panel:} Same, overlaid on the galaxy density contours computed from the photometric redshift catalogue by \cite{2022yCat.7292....0D}.}
\label{fig:image1330}
\end{figure}

The XCLASS~1330 galaxy structure located at RA $ = 14^{\rm h}\ 08^{\rm m}\ 26^{\rm s}$, DEC $ = 78^\circ \ 36^\prime\ 30^{\prime\prime}$ (J2000), based on the X-ray emission peak, is part of the XCLASS survey. Its redshift is $z\sim0.19$.
It has a complex structure with at least four different SbS and we were not able to run the FIREFLY and PIPES\_VIS codes due to the lack of SDSS spectra. 
XCLASS~1330 has the peculiar property of showing bright galaxies in non-central locations (e.g. structure 111 in Fig.~\ref{fig:image1330}), rather than in its inner X-ray emitting core (structure 12 in Fig.~\ref{fig:image1330} at the same redshift). It is also visually complex and will prove to be dynamically building up, as is discussed below.

Since XCLASS~1330 is located at a high declination, out of reach of most ground-based telescopes, only a few galaxies have public spectroscopic redshifts in NED (NASA/IPAC extragalactic database): WISEA J140915.24+783937.2 ($z=0.0901$), NGC 5547 ($z=0.0390$), and NGC 5452 ($z=0.0069$), all of these being foreground galaxies. This makes its vicinity very poorly known in terms of cosmological nodes and massive clusters.
Taking advantage of the relatively northern location of the MISTRAL spectrograph at OHP\footnote{http://www.obs-hp.fr/welcome.shtml: OHP is at latitude $+43^\circ56^\prime$}, we measured 28 supplementary redshifts 
for galaxies brighter than $r$'$\simeq$20 mag (see Tab.~\ref{tab:NewMISTRAL}). They were selected within a radius of $4^\prime$ from the XCLASS~1330 BGG X-ray peak, corresponding to a physical radius of $\sim$800~kpc at the group redshift. This is enough to sample the group itself plus its vicinity, including potential infalling or neighbouring  structures.

\begin{figure}[ht]
\begin{center}
\includegraphics[width=0.4\textwidth]{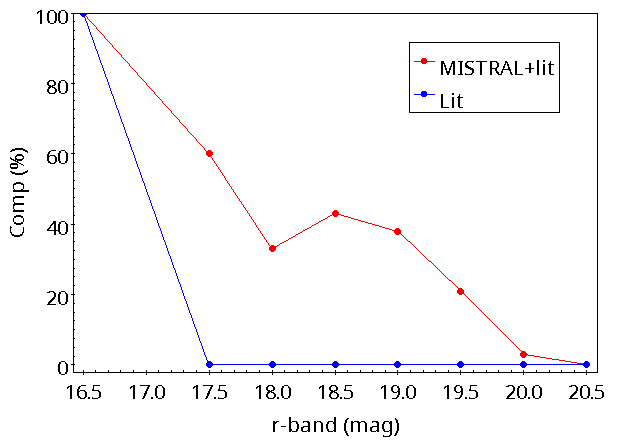}
\end{center}
\caption{Completeness of our redshift catalogue along the XCLASS~1330 line of sight as a function of r-band magnitude. The blue line represents literature data; the red line MISTRAL+literature data.}
\label{fig:comp1330}
\end{figure}

In order to estimate our spectroscopic catalogue completeness, we selected all galaxies from the Pan-STARRS database (applying $\|$rpsfLikelihood$\|\leq 0.001$ as the condition\footnote{rpsfLikelihood is the Pan-STARRS criterion allowing the star-galaxy separation based on r-band data.} to remove stars from the sample) within the sampled radius and compared them with our spectroscopic catalogue. We clearly see in Fig.~\ref{fig:comp1330} that the completeness of our new spectroscopic catalogue is close to 40--50$\%$ up to r$\sim$19, and then drops down rapidly to very low values.  Given the distance modulus of XCLASS~1330, this corresponds to a sampling of the group down to absolute magnitudes of ${\rm M_r} \sim -21$, close to the M$^*$ value at the redshift of the considered group \citep[see e.g.][]{2005A&A...439..863I}.

In order to compensate for this incompleteness, we also considered the photometric redshift catalogue along the XCLASS~1330 line of sight from \cite{2022yCat.7292....0D}. We generated Fig.~\ref{fig:image1330} (lower figure) by selecting photometric redshifts between 0.05 and 0.30 down to r=20 and computing density contours with the topcat\footnote{http://www.starlink.ac.uk/topcat/} software. This figure shows galaxy concentrations well correlated with the SbS detected by the SG software, in particular matching the 12 and 111 structures very well (see below). 

We have neither SDSS nor 6dFGS spectroscopy for XCLASS~1330, so we could not perform spectral energy
distribution modelling, as the MISTRAL spectroscopy was only designed to measure redshifts (no flux calibration, signal-to-noise (S/N) only adapted 
to identify the lines, and a spectral resolution of R$\sim$700). However, we used the SG tool to detect potential SbS.

We detect the inner core (structure 12, pink ellipse in Fig.~\ref{fig:image1330}) of the group as an independent low-mass (only upper-constrained mass by the SG code) structure made of four galaxies at $z \sim 0.1972$, and with a short crossing time (only upper-constrained as well). This internal structure matches the XMM-Newton X-ray contours very well (see Fig.~\ref{fig:image1330}) and probably has a very high merging rate.
We note that XMM-Newton data are not deep enough for the mass to be directly computed. Other structures are also present:
 
 -- Structure 11 (at $z\sim 0.1955$) is made of two SbS: a structure labelled 111, with six galaxies (see Fig.~\ref{fig:image1330}), and, if we diminish the number of galaxies per structure to two, structure 112 (blue ellipse
in Fig.~\ref{fig:image1330}). Structure 111 is more massive than structure 12 (these are only upper-constrained SG-estimated masses) and has a similarly short crossing time, implying a high merging rate. We do not know if it has an X-ray counterpart because XMM-Newton data do not cover the full field of view (only structures 2 and 12 are covered).

-- Structure 2 (grey ellipse in Fig.~\ref{fig:image1330}), at a similar redshift of 0.1955, has a low SG-estimated mass and a short crossing time. Its central part also corresponds to an XMM-Newton X-ray emission.

It is therefore likely that we are dealing with a structure that is still building up, probably typical of the merging of low-mass groups in a filament of galaxies. 

\section{The MCG+00-27-023 massive group}\label{sec:MCG+00}

\begin{figure}[ht]
\begin{center}
\includegraphics[width=0.4\textwidth]{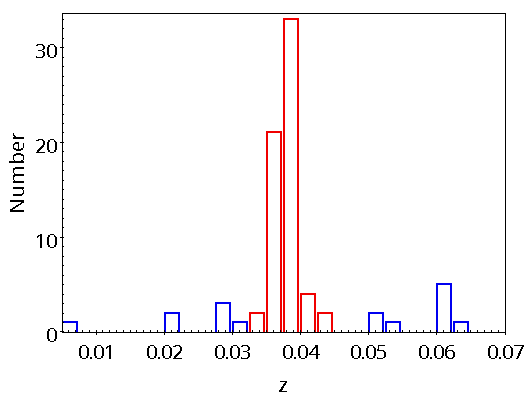}
\includegraphics[width=0.4\textwidth]{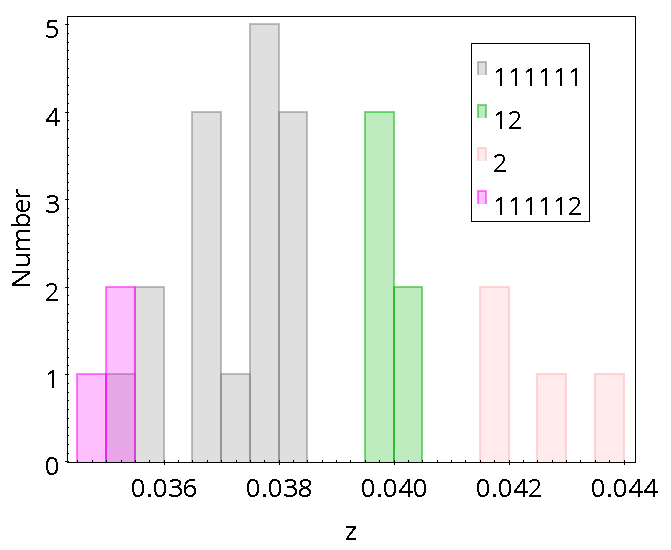}
\end{center}
\caption{Upper panel: Redshift histogram of galaxies in the field of MCG+00-27-023. Lower panel: Redshift histogram of galaxies in dynamically linked 
structures with more than three members within the [0.034, 0.044] redshift interval.}
\label{fig:histozMCG}
\end{figure}

\begin{figure}[ht]
\begin{center}
\includegraphics[width=0.4\textwidth]{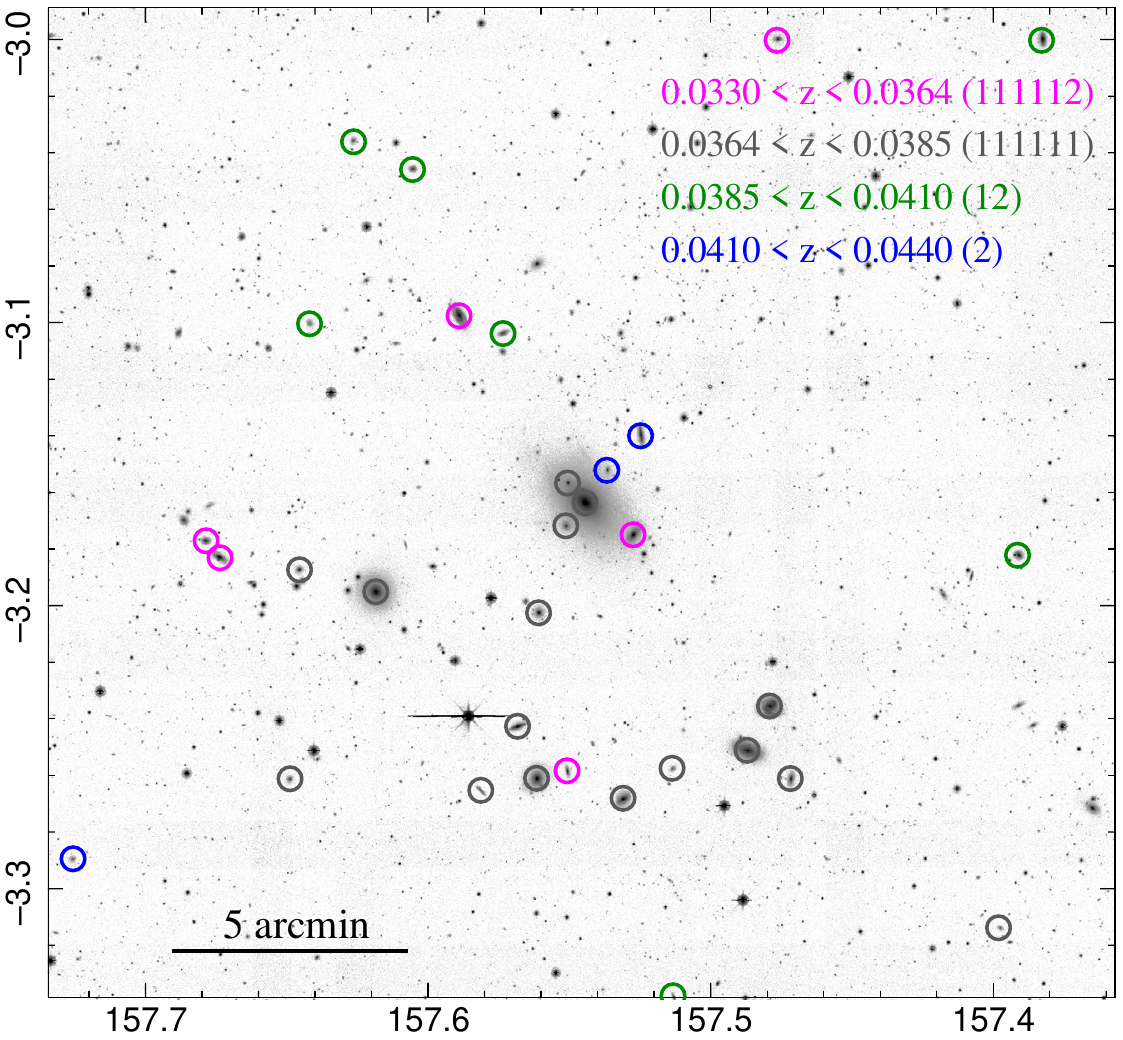}
\end{center}
\caption{Sky distribution of the galaxies belonging to the MCG+00-27-023 line of sight within the redshift interval [0.033; 0.044]. Colours are related to the different detected Sbs: pink, group 111112; grey, group 111111; green, group 12; and blue, Group2. }
\label{fig:RADECMCGnew}
\end{figure}

The MCG+00-27-023 group is located at RA $ = 10^{\rm h}\ 30^{\rm m}\ 10.7^{\rm s}$, DEC $ = -03^\circ \ 09^\prime\ 49.1^{\prime\prime}$ (J2000), and its redshift is $z=0.03725$.  
It was already detected as a relatively massive structure by 
\cite{KoranyiGeller02}. It is known to be made of two components (MKW2a and b), and therefore fits our search for groups undergoing a building phase.

To find a more massive structure in terms of velocity dispersion in the vicinity of MCG+00-27-023, we searched the C4 SDSS catalogue \citep{Miller+05}. The two most massive clusters known within a 35~Mpc radius are C4-1110, with a 
velocity dispersion of 405 km~s$^{-1}$ at 24~Mpc, and C4-1023, with a clearly larger velocity dispersion of 1247 km~s$^{-1}$, and distant from MCG+00-27-023 by 36 Mpc. This last cluster is also known as Abell~0957 \citep[see e.g.][]{Mazure+96}. 
Such a massive cluster is probably in the middle of a cosmological node given its high dynamical mass. We therefore assume that the closest node from  MCG+00-27-023 is located at $\sim$36 Mpc.

We selected NED galaxies with a known spectroscopic redshift between $z=0.005$ and 0.07 and within a radius of $\sim$700~kpc from the BGG. The redshift histogram for these galaxies in the MCG+00-27-023 field of view is shown in Fig.~\ref{fig:histozMCG}. 
A clear peak appears between 0.034 and 0.044. We show details of this peak in Fig.~\ref{fig:histozMCG} (lower graph).
This main system has a velocity dispersion of 565 km~s$^{-1}$ and a mean redshift of
0.0369. FIREFLY measures a main stellar population age of 6.8 Gyrs (after the Big Bang, hereafter aBB)
and a metallicity of 0.7 Z$_{\odot}$. The 
spectrum of this stellar population is best modelled by FIREFLY
with four bursts (see Table.~\ref{tab:firefly}): a dominant burst (in terms of mass, nearly 50$\%$ of the created stellar
mass) 10 Gyrs aBB, two other bursts 4 and 2 Gyrs aBB, and a very recent (11 Gyrs aBB) low-mass one (less
than 5$\%$ of the total stellar mass).

We then applied the SG dendrogram tool to our sample of galaxies. Considering only structures with more than three galaxies, we detect two independent subgroups (structures 2 and 12) at slightly different redshifts than that of MCG+00-27-023. Group 12 is a small group of six galaxies at z$\sim$0.0391, with an estimated 
virial mass of $0.3 \times 10^{13}$~M$_{\odot}$. The latest burst
of star formation in structure 12 is predicted by FIREFLY to have occurred 12 Gyrs aBB. This age probably corresponds to the merging time of structure 12 with MCG+00-27-23 itself. 
Group 2 is a more massive structure at $z \sim 0.0416$ sampled with four galaxies and an estimated virial mass of
$3.9 \times 10^{13}$~M$_{\odot}$ (but its virialized state is not clear because it is sampled by a relatively small number of galaxies). 

MCG+00-27-023 appears as the 111 structure, sampled with 42 galaxies at $z\sim 0.0365$. It has an estimated virial mass of $2.3 \times 10^{13}$ M$_{\odot}$. 
We detect in the 111 structure a single (minor) SbS with more than three galaxies: 111112, with a mean redshift of 0.0345 and a SG-estimated virial mass $\leq 0.1 \times 10^{13}$~M$_{\odot}$ 
(see Fig.~\ref{fig:RADECMCGnew}). It probably constitutes a relatively recent structure that has merged with MCG+00-27-023. The latest burst
of star formation in SbS 111112 is predicted by FIREFLY to have occurred 13 Gyrs aBB, refined to $\sim$11.5 Gyrs with \texttt{PIPES\_VIS} to explain the strength and ratio of the H$\alpha$,
[OIII] and H$\beta$ lines. This age probably corresponds to the merging times of SbS 111112 with MCG+00-27-023 itself.

The 111 structure is dominated by a pair of galaxies (MCG+00-27-023 itself and LEDA 30995). These are the brightest and second-brightest galaxies in the group, with SDSS $r$-band magnitudes of $\sim$12.66 mag and $\sim$14.12 mag, respectively, resulting in a magnitude gap of $\Delta m_{12} = 1.46$. Therefore, the group is not considered a fossil system. The $r$-band magnitude of the third galaxy is $\sim$14.33 mag. However, these magnitudes do not take into account the extended stellar envelopes of the galaxies, as was discussed for the NGC~4104 group (see Sect.~\ref{sec:N4104} below). For this reason, we also performed a detailed 2D decomposition of the galaxies to include their extended stellar envelopes in the photometry (see Appendix~\ref{ap:decomp}).

The structure we denote as 1111111 can be considered as the inner core of the MCG+00-27-023 group. Made of eight galaxies, it has
a SG-estimated velocity dispersion of 174 km~s$^{-1}$ and a virial mass of $0.3 \times 10^{13}$ M$_{\odot}$. This structure
has an E(B-V) close to that of the main system (0.219), a slightly younger mean stellar population (formed 10.5 Gyrs
aBB), and a comparable metallicity of 0.6~Z$_{\odot}$. This is also visible when considering the bursts
required by FIREFLY to model the summed spectrum: three
bursts between 10 and 12 Gyrs aBB. This means that star formation was active inside this structure until 1 Gyr
in the past as counted from today.  However, lines as H$\alpha$ or [OIII] have now vanished,
indicating that the star formation activity is presently at most very weak.
We note that we still see the [NII]6583 emission line in the summed spectrum of structure 1111111, but this is
probably due to the central black hole activity
\citep[e.g.][]{2023AdSpR..71.1219D}.

The SG-estimated crossing time can give an estimation of the time before the next galaxy-galaxy merger inside the
1111111 structure: 1.3 Gyrs. Galaxy-galaxy mergers can occur if there is energy exchange between the galaxies to decrease their relative velocity. These energy exchanges occur more efficiently during (slow) close-encounters \citep{2000ASPC..197..377M}.

The MCG+00-27-023 system therefore corresponds to the classical group representation: a low-mass
structure of a few 10$^{12}$~M$_{\odot}$, with regular infalling events along its history and a relatively low
value of the mean metallicity in the core structure of 
0.6~Z$_{\odot}$. The crossing time of the inner core
structure (1111111) is also large enough (1.3 Gyrs) to prevent any total merging of the member galaxies
from happening for a relatively long time. The MCG+00-27-023 system is therefore very different from FGs, where this
total merger has probably already occurred.

\section{The NGC~4065 non-fossil group}
\label{sec:UZC}

The NGC~4065 group is located at RA $ = 12^{\rm h}\ 04^{\rm m}\ 06.17^{\rm s}$, DEC $ = +20^\circ \ 14^\prime\ 06.4^{\prime\prime}$ (J2000), and its redshift is $z=0.02106$.  
This group has a complex dynamical structure 
\citep[e.g.][]{1999AJ....118.2014W}
 and is located within the same cosmic bubble as NGC~4104 (see Sect.~\ref{sec:N4104}), but closer to the dominant cluster of the area, the Coma cluster, only at 13 Mpc (3D). With a mass close to 10$^{15}$~M$_\odot$, Coma largely dominates the NGC~4065 group in terms of mass (see below) and can be considered as the closest cosmological node. 

\subsection{SG and FIREFLY combined analysis}
Similarly to MCG+00-27-023, we applied the SG tool to our sample of galaxies of the NGC~4065 group considering structures with more than three galaxies (see Figs.~\ref{fig:histoz4065} and \ref{fig:RADEC4065}). We detect two independent structures (112 and 1112). Structure 112 is a small group of three galaxies at $z\sim 0.0228$, with a SG-estimated 
virial mass $\leq$ $0.1 \times 10^{13}$~M$_{\odot}$. 
Group 1112 is a
similar structure at $z\sim 0.0243$ sampled with six galaxies 
and with an estimated virial mass of $\leq$ $0.1 \times 10^{13}$~M$_{\odot}$. They have 
probably merged very recently with the main NGC~4065 group: the latest bursts
of star formation in these two SbS are predicted by FIREFLY to have occurred 13 Gyrs aBB.
The use of \texttt{PIPES\_VIS} refines the value to $\geq$11 Gyrs for structure 1112 to explain the strength and ratio
of the H$\alpha$, [OIII], and H$\beta$ lines. These ages probably correspond to the merging times of
structures 112 and 1112 with the main NGC~4065 group.

The inner 
core of the NGC~4065 group itself appears as the 1111 structure, sampled with 12 galaxies at $z\sim 0.0226$. It has a low SG-estimated virial mass of $\leq$ $0.1 \times 10^{13}$~M$_{\odot}$. FIREFLY measures
a main stellar population age of 5.9 Gyrs (aBB) and a metallicity of 1.3 Z$_{\odot}$. The summed spectrum of
this population is best modelled by FIREFLY with five bursts, including a very minor one accounting
for less than 1$\%$ of the total stellar mass generation (see Table~\ref{tab:firefly}). There is no dominant burst among the four main ones.
The earliest occurred very soon in the history of the Universe, another one 4~Gyrs aBB, and the two last ones 9~Gyrs aBB (FIREFLY probably split a single burst into two entities). 
This means that star formation was not very active recently inside this structure, and indeed
H$\alpha$ and [SII] are barely visible. The only very significant emission line is [NII]6583, but
it is probably not related to any star formation. 
The SG-estimated crossing time can give an order of the time before the next galaxy-galaxy merger inside the
1111 structure: $\leq$0.1 $\times$ 10$^{9}$ years. This probably induces frequent galaxy-galaxy mergers. 
The case of the NGC~4098 galaxy (member of the 1112 structure, recently merged with the NGC~4065 main structure) is a good example of these mergers. In order to better understand how this galaxy produced its large tidal tails, we observed it in integral field spectroscopy. 

\subsection{Integral field spectroscopy of NGC~4098}

We observed NGC~4098 with the Fabry-Perot instrument GHASP 
used on the OHP 193 cm telescope. GHASP has a field of view 
of 5.8 × 5.8 arcmin$^2$, and is coupled with a 512 × 512 
image photon counting system with a pixel size of 
0.68 $\times$ 0.68 arcsec$^2$. Observations were made with a  configuration similar to that already described in \cite{2023A&A...680A.100A}. The data reduction procedure adopted to reduce the GHASP data is extensively
described in \cite{2019A&A...631A..71G}.
We also present in Appendix \ref{ap:ngc4098} details specific to this galaxy system, including a $g$, $r$, $z$-band colour mosaic of NGC 4098, reproduced from the Siena Galaxy Atlas 2020 \citep{2023ApJS..269....3M} (see Fig. \ref{fig:NGC4098-largegalaxy-grz-montage}). 
The first two images show that the system is made up of two galaxies separated by $\sim$5.1~kpc ($\sim$0.93~arcsec). Measured from the bottom left panel, the z-band radius is R$_{z26}\sim 64$ arcsec (represented by the semi-major axis of the blue ellipses in the bottom line panels), leading to a diameter for the system of D$_{z26}\sim$65~kpc, much too large to be a single galaxy.

\begin{figure*}[ht]
    \centering
    \includegraphics[width=\textwidth]{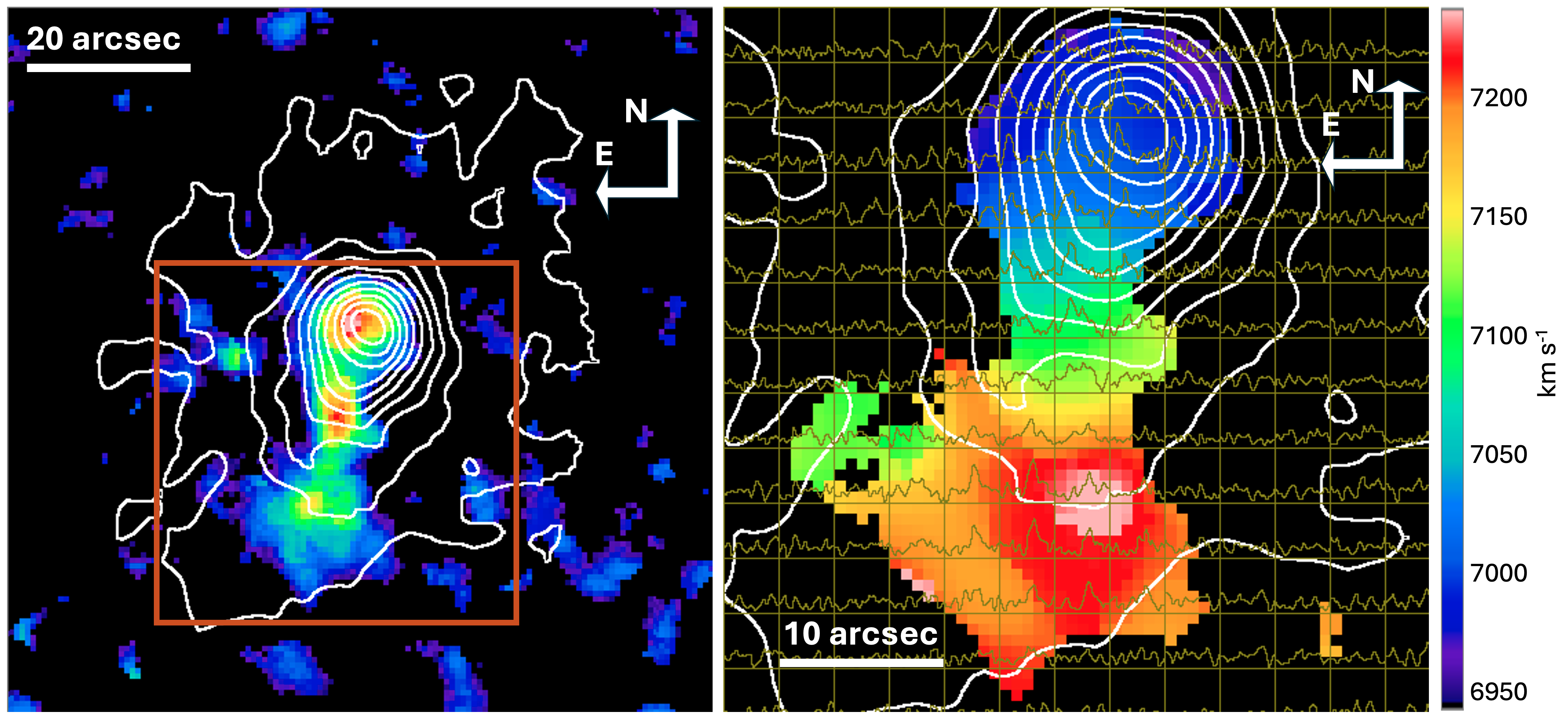} 
    \caption{%
        \textbf{Left:} H$\alpha$ emission line map of NGC 4098 and 4099, on which is superimposed the continuum emission below and beside the H$\alpha$ line within a FWHM of 15~\AA. The size of the total region shown is $\sim$1.45$\times$1.45 arcmin$^2$ ($\sim$44.1 $\times$ 44.1 kpc$^2$). Flux is in arbitrary units. 
        \textbf{Right:} H$\alpha$ line-of-sight velocity field zoomed by a factor of 2 from the brown square drawn on the left panel, on which are overplotted the same continuum isocontours. The size of the total region shown is $\sim$43.5 $\times$ 43.5 arcsec$^2$ ($\sim$22.1 $\times$ 22.1 kpc$^2$). Each box containing each profile has dimensions of $\sim$3.4 $\times$ 3.4 arcsec$^2$ ($\sim$1.7 $\times$1.7 pc$^2$), which correspond to a binning of 5 $\times$ 5 pixels; in other words, each box is the profile average of 25 spaxels. The x scale of each box spans over the free spectral range of the interferometer, ranging from $\sim$6896 to $\sim$7376 km s$^{-1}$, the sampling step being 10.25 km s$^{-1}$. The y scale is given in arbitrary units. These data have been observed with GHASP on the T193 at OHP.
    }
    \label{fig:ngc4098_VF}
\end{figure*}

\begin{figure}[ht]
\begin{center}
\includegraphics[width=0.44\textwidth]{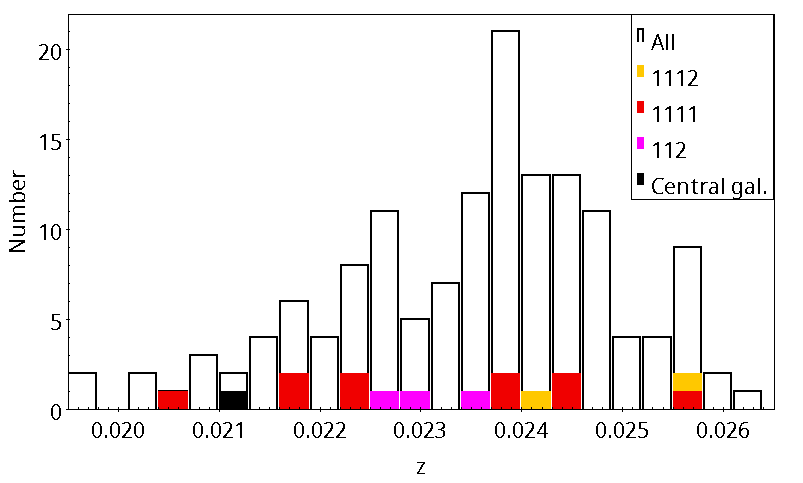}
\end{center}
\caption{Redshift histogram of the galaxies in the field of the NGC~4065 group.}
\label{fig:histoz4065}
\end{figure}

\begin{figure}[ht]
\begin{center}
\includegraphics[width=0.4\textwidth]{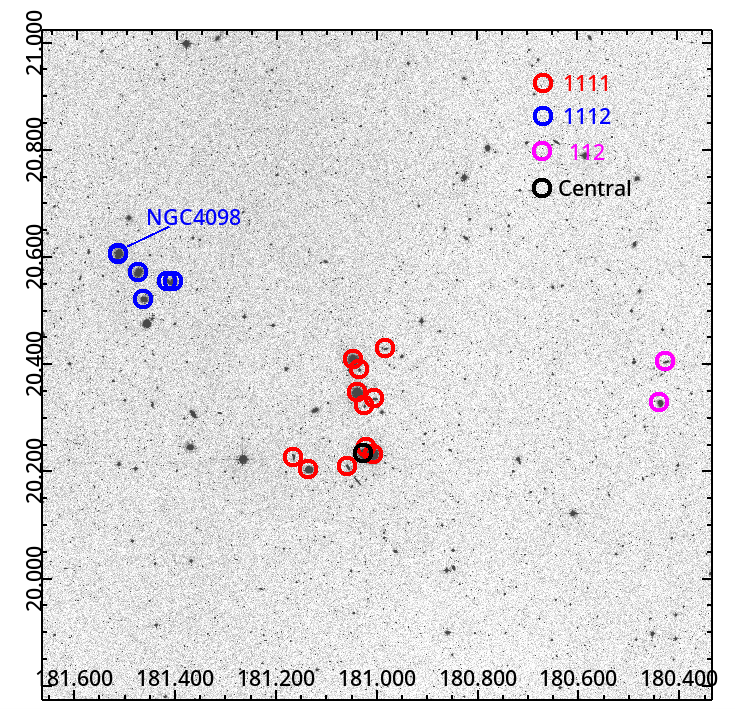}
\end{center}
\caption{Sky distribution of the galaxies belonging to the NGC~4065 group.}
\label{fig:RADEC4065}
\end{figure}

The emission from the $g$-band image is displayed in light blue in the first two panels at the top row of Fig. \ref{fig:NGC4098-largegalaxy-grz-montage}. It traces recent star formation around the galaxy system, in a ring structure $\sim$28 kpc ($\sim$55 arcsec) in diameter, whose major axis position angle is about 69 degrees (from north) and inclined about $\sim$45$^\circ$ on the plane of the sky. This ring-shaped structure shows multiple pieces of tidal debris, tracing a recent galaxy orbit that left debris in its wake. Following the debris, one can imagine seeing the trace of two successive orbits.  No H$\alpha$ counterpart is detected along this blue ring (see Fig. \ref{fig:ngc4098_VF}). A dust lane is observed between the northern and southern lobes; this lane corresponds to where the H$\alpha$ emission is very low, probably extinguished by dust at the top left image. In the GALEX image (second image of the bottom panel of Fig. \ref{fig:NGC4098-largegalaxy-grz-montage}), whose resolution is low, the UV counterparts of this ring visible in the g-band image can nonetheless be discerned, suggesting that the stars in this ring are between 10 and 100 Myrs old.

Older debris is also observed in the r- and z-band images of the top panels of Fig. \ref{fig:NGC4098-largegalaxy-grz-montage}, as is underlined in the rightmost residuals image, which shows the difference between the data and the model (the two leftmost images). This older debris lies along a north-south axis passing through the NGC~4098 centre. The structure to the north extending over $\sim$20 kpc ($\sim$40 arcsec) is particularly knife-cut on the west side.  The WISE W1/W2 image (third panel of the bottom line) shows that stellar emission is found at large radius, at least up to the red ellipse. In the same image, one can hardly distinguish an enhancement of stellar emission around NGC 4099. The WISE 12 micron emission (rightmost bottom line panel) shows the dust emission as a relic of the ancient debris of previous interactions.

The left panel of Fig. \ref{fig:ngc4098_VF} displays, within the extended continuum isocontours, the net H$\alpha$ emission of the system, which is more limited in size. The shape, size and inclination of $\sim$45$^\circ$ of the continuum map  around H$\alpha$ agree with the quantities estimated from broad band images, and probably trace the debris of the oldest stars due to the interaction between the two galaxies.  The northernmost H$\alpha$ emission corresponds to the peak continuum flux from the galaxy NGC 4098, slightly shifted eastwards. The central emission aligns with galaxy NGC 4099, where continuum isocontours reveal an underlying stellar component. Importantly, the southernmost emission shows no visible stellar counterpart in the continuum map, indicating that no continuum emission is seen south of galaxy NGC 4099.

The right panel of Fig. \ref{fig:ngc4098_VF} displays a regular line-of-sight velocity gradient from north to south along the velocity field, with a total amplitude of $\sim$300 km~s$^{-1}$, spanning over $\sim$22 kpc.  This suggests a single kinematic structure.  Within the main body of NGC 4098 extending over $\sim$7.5 kpc, a small velocity gradient of about 5 km~s$^{-1}$ kpc$^{-1}$ is observed. We measure a disk inclination of 57$^\circ$ giving a maximum rotation velocity amplitude of $V_{max}\sim$21~km~s$^{-1}$, with a corresponding mass of $\sim4 \times 10^{8}$ M$_\odot$. However, this result is highly sensitive to the galaxy inclination, and the Tully-Fisher relation indicates that the mass of a galaxy of this size could be $\sim9.5 \times 10^8$ M$_\odot$. Thus, with an inclination of $\sim$10$^\circ$ instead of 57$^\circ$, then $V_{max}\sim$100 km~s$^{-1}$, and the mass estimate would rise by more than a factor of 20.  The large overall gradient of $\sim$300 km~s$^{-1}$ could be due to streaming motions due to the interaction between galaxies, and form a 3D pattern as is suggested by the velocity peak at $\sim$7243 km s$^{-1}$, which is not located at the end of the tail. The velocity difference between the peak velocity of the tail and NGC 4098 is approximately 
$\sim$220~km~s$^{-1}$ over a distance of $\sim$11 kpc, implying an age for the tail of $\sim$50 Myr.

A detailed examination of the H$\alpha$ profiles, illustrated in the left panel of Fig. \ref{fig:ngc4098_VF}  where 25 pixel boxes have been integrated, shows complex and multiple structures, with at least two major components. It is clearly seen that the main peak of the profile is progressively redshifted from north to south. We observe a continuity between the two galaxies and note that the two main components have almost the same H$\alpha$ flux.  Figure \ref{fig:ngc4098_VelocityProfiles} complements Fig.  \ref{fig:ngc4098_VF} and clearly shows the two components, mainly the ‘blue’ and ‘red’ components.  The integrated H$\alpha$ profile over the entire system matches well with the HI integrated profile, the latter nevertheless being more extended at large velocities.

The observations of NGC 4098 can be summarized as follows:
(1) Broad-band images reveal both young and old stellar populations, multiple plumes of varying ages, and dust lanes, all enveloped within an extensive structure exhibiting an $r^{1/4}$ surface brightness radial profile.
This may be an indication of an old relaxed structure (coming from the 4 Gyrs aBB event detected in SbS 1112, see Table B.1), on top of which the most recent burst of SbS 1112 is detected.
(2) The H$\alpha$ distribution is asymmetric relative to the galaxy centre and extends southwards, with three aligned H$\alpha$ regions.
(3) The southernmost H$\alpha$ blob lacks a stellar counterpart, while the central H$\alpha$ blob, associated with NGC 4099, has a faint stellar component.
(4) A continuous velocity gradient is observed across the velocity field from north to south, except for a localized velocity jump in the southern region.
(5) There is agreement between the velocity and flux distributions in both the cold (HI) and warm (H$\alpha$) hydrogen components.
(6) The H$\alpha$ profiles exhibit multiple complex components, dominated by two lines of similar intensity, potentially representing two different components observed due to projection effects along the line of sight.
These observations suggest the presence of an extended tidal tail curving in a plane perpendicular to the sky. The two southern H$\alpha$ blobs, including NGC~4099, may be tidal dwarf galaxies (TDGs), similar to those observed in many interacting systems where gravitational clumps form along tidal tails.  VV062 and NGC 4098 (i.e. VV061) are separated by $\sim$105~kpc ($\sim$208 arcsec). The east-west HI bridge linking the galaxies extends over $\sim$165 kpc, and even up to $\sim$196~kpc if measured from the longest extension from east to north-west; thus, VV062 could be a past intruder.

\subsection{Global behaviour of the NGC~4065 group}

The NGC~4065 system is not a typical group. It has a low mass but its metallicity is already relatively high as compared to
MCG+00-27-023: 1.3 Z$_{\odot}$, and the crossing time of the system is also very short. These quantities indicate
an intense merging activity leading to a high metal diffusion into the system. The crossing time of the inner
core structure (1111) also makes a full merger of the member galaxies possible
in a short time (a fraction of a gigayear). The NGC~4065 system is therefore possibly following the path to become a FG
within a relatively short time, in the framework of the total fusion model of FG generation.

However, we must note that if we consider the global structure (including the previously detected substuctures), the global estimated mass is of the order of $4 \times  10^{13}$~M$_{\odot}$.
The large difference between this total mass and the sums of the individual masses of the various SbS is probably an indication of the incomplete virialization state of the parent structure. As was mentioned before, SG mass estimates are therefore to be considered with caution.

\section{The NGC~4104 near-FG}
\label{sec:N4104}

NGC~4104 is the dominant galaxy of a group recently studied by \cite{LimaNeto+20}. It is located at RA$ = 12^{\rm h}\ 06^{\rm m}\ 38.9098^{\rm s}$, DEC $ = +28^\circ \ 10^\prime\ 27.18^{\prime\prime}$ (J2000), and its redshift is $z=0.0286$.
This structure proved to have a real merging history with the visible result being concentric residual stellar shells.
Its cosmic bubble is dominated by the massive clusters Abell~1367 (at $\sim$~32~Mpc 3D) and Coma (at $\sim$~29~Mpc 3D). Other groups are present such as the NGC~4065 (see Sect.~\ref{sec:UZC}) and NGC~4295 groups, but only with moderate masses untypical of massive clusters. NGC~4104 is therefore relatively far from any cosmic node.

\begin{figure}[ht]
\begin{center}
\includegraphics[width=0.44\textwidth]{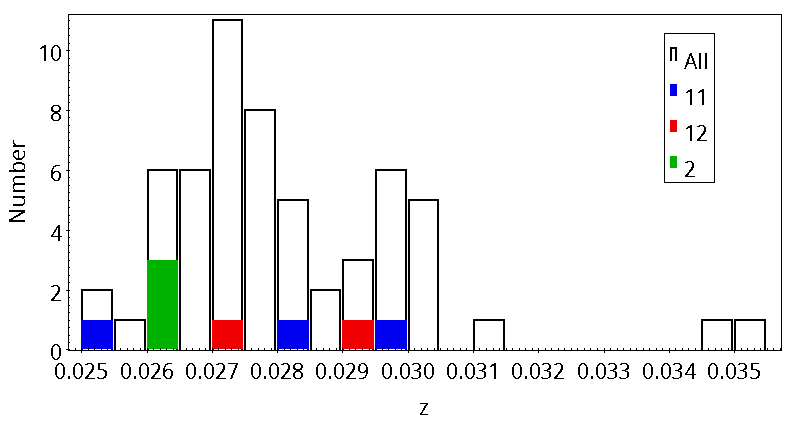}
\end{center}
\caption{Redshift histogram of the galaxies in the field of the NGC~4104 group.}
\label{fig:histoz4104}
\end{figure}

We already presented the NGC~4104 group in a previous work  \citep{Chu+23}, and re-analysed this structure with an updated galaxy redshift catalogue. 
This redshift catalogue was collected in the NED\footnote{https://ned.ipac.caltech.edu/} database within a 60~arcmin radius around the position of NGC~4104, limited to the [0.01, 0.05] redshift range (see Fig.~\ref{fig:histoz4104}). We only considered spectroscopic redshifts. 
We then performed the same SG SbS search as for the other  galaxy structures considered in the present paper.

Within the main bulk at z$\sim$0.028 (with an SG-estimated mass of $1.4 \times 10^{13}$M$_{\odot}$), FIREFLY measures a main stellar population age of 8.5 Gyrs (aBB)
and a high metallicity of 1.8 Z$_{\odot}$. The summed spectrum of this population is best modelled by FIREFLY
with three bursts (see Table~\ref{tab:firefly}): a dominant one (in terms of mass, nearly 50$\%$ of the created stellar
mass) 10 Gyrs aBB, and two other bursts 2 and 12 Gyrs aBB (the latter adjusted to 10.7 Gyrs aBB to explain the total absence
of H$\alpha$ and [SII], and the small [NII] emission).

The NGC~4104 group presents a small amount of SbS. 
Only two subgroups are detected by the SG method. The first one (11) with three galaxies, including NGC~4104 itself, has an estimated mass of $7.1 \times 10^{12}$~M$_{\odot}$ and represents the core of the group.
It is typical of a massive group, and its crossing time is of the order of 0.4 Gyr, still quite short compared
to what happens in MCG+00-27-023. FIREFLY models the summed spectrum with five bursts: 1.5, 4, 5, 10, and 12 Gyrs aBB
with no dominant burst. The age of the most recent one is adjusted by PIPES\_VIS to 11~Gyrs to explain the [NII]/H$\alpha$ ratio. 

\begin{figure}[ht]
\begin{center}
\includegraphics[width=0.4\textwidth]{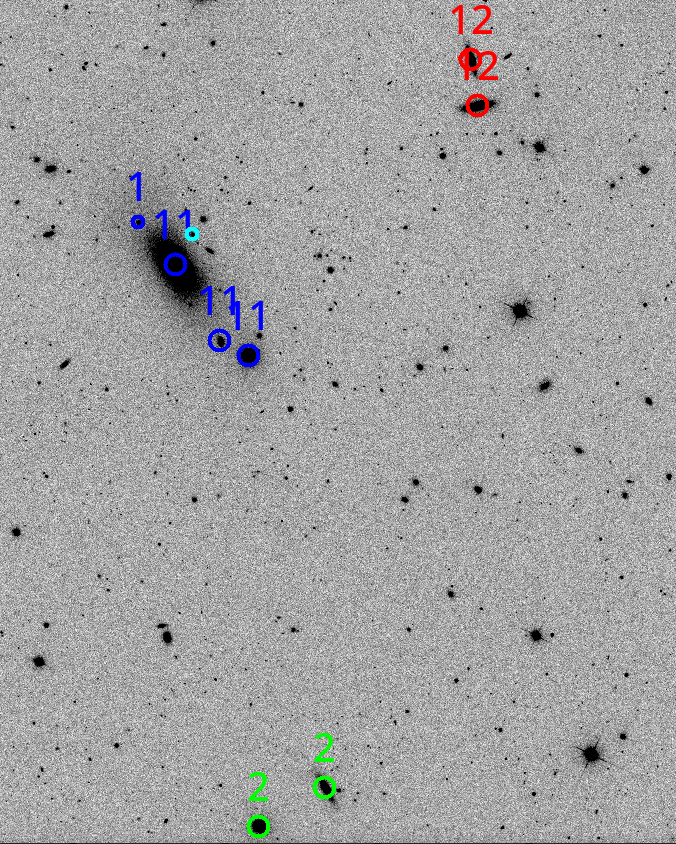}
\includegraphics[width=0.4\textwidth]{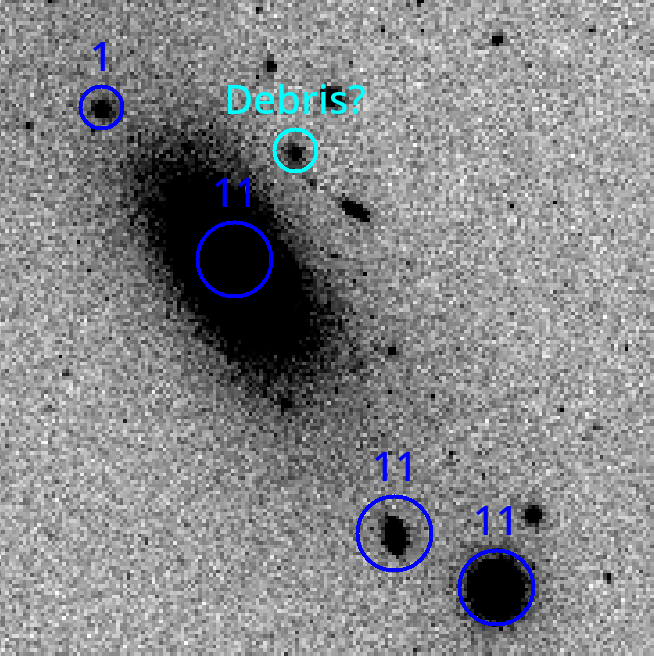}
\end{center}
\caption{Sky distribution of the galaxies belonging to the SbS detected within the NGC~4104 group. The bottom panel is a zoom-in of the central galaxy.}
\label{fig:RADEC4104}
\end{figure}

The other SbS (12) is made of two 
galaxies north-west of NGC~4104 (see Fig.~\ref{fig:RADEC4104}) and is probably falling into the core of the group.
It has an estimated mass of $5.3 \times 10^{12}$ M$_{\odot}$. The last burst of star formation is detected by FIREFLY at 12 Gyrs aBB. The adjustment by PIPES\_VIS leads to 10.5~Gyrs for structure 12 to explain the complete absence of H$\alpha$ emission. 

We finally detect another minor structure (number 2 in Fig.~\ref{fig:RADEC4104}), located 500 km~s$^{-1}$ ahead
of NGC~4104, so probably weakly interacting with the core of the group.
It has an estimated mass of $\leq 0.1 \times 10^{13}$ M$_{\odot}$. The last burst of star formation is detected by FIREFLY 13 Gyrs aBB, adjusted to 12~Gyrs by
PIPES\_VIS in order to explain the small H$\alpha$ emission (of the order of [NII]), without any [OIII] emission. 

The NGC~4104 system is not typical of a classical group, since it already has a relatively high mass
of the order of $7 \times 10^{12}$~M$_{\odot}$ and a high metallicity of 1.8 Z$_{\odot}$. The crossing time of the system is also short: 0.5 Gyrs. These two values
 indicate, as for the NGC~4065 system, a likely intense merging activity leading to a large metal
diffusion into the system. 

Strictly speaking, the NGC~4104 group can only be classified as a FG if we take into account the extended stellar envelope of the central galaxy. If we consider only the central component, without the extended envelope, it cannot be classified as a FG because the magnitude gap between the first and second galaxies would be only 1.4~mag in the r-band (see also 
\citep{LimaNeto+20}). Even considering the group inner core (SbS 11 in the SG analysis), this magnitude gap does not reach two magnitudes (1.66).  However, the NGC~4104 galaxy is still largely dominant within the group (see e.g. Fig.~\ref{fig:RADEC4104}) and 
the merging process could soon make NGC~4104 a real FG given the relatively short crossing time estimated by the SG method within structure 11: less than a gigayear.

In this framework, we also have proof of active gravitational interactions close to NGC~4104. We detected a likely piece of debris, part of structure 1, which could also be part of the inner core group (structure 11), following the SG analysis (with a probability of 54\%). 
This debris is shown in Fig.~\ref{fig:RADEC4104} and has a disturbed shape. It also has an emission line spectrum, with a clear H$\alpha$ line. 

\section{Candidate fossil groups in the CFHTLS} 
\label{sec:CFHTLS}

Following the detection of 1371 candidate clusters and groups in the  CFHTLS based on galaxy
photometric redshifts by \cite{Sarron+18}, candidate FGs were selected statistically 
by \cite{Adami+20}, applying the condition
that the second brightest galaxy should be at least two magnitudes fainter than the brightest one within the R$_{500}$ radius. For all the galaxies, we only had photometric redshifts, so in order to confirm that these were indeed candidate FGs, we searched various databases for spectroscopic redshifts for the BGG of each FG and found a few.  For a number of groups, we then obtained spectroscopy with various telescopes for several of the brightest galaxies in the region covered by each group, to measure galaxy redshifts. 
We must note, however, that even in the case of an optical confirmation, these objects remain candidate FGs in the absence of X-ray data.

\subsection{Spectroscopic observations}

Long-slit spectroscopy was obtained with MISTRAL at Observatoire de Haute-Provence, with ALFOSC at the Nordic Optical Telescope, and with CAFOS at the Calar Alto Hispano Alem\'an Telescope. Multi-object spectroscopy was also obtained (mainly for the faintest objects) at the Telescopio Nazionale Galileo with DOLORES in November of 2023.

The list of galaxy spectroscopic redshifts is given in Table~\ref{tab:spectro} for the 14 candidate FGs.  For each group, the first line indicates the group central co-ordinates taken from the \cite{Sarron+18} catalogue and the spectroscopic redshift of the BGG. The following lines show the galaxies in order of increasing $r$ band magnitude. When no redshift is given, this could be due to either the lack of observations or the weakness of the spectrum, which prevents us from measuring a reliable redshift.

With these spectroscopic data, only two candidate FGs remain FG candidates in the optical: W3\_FG65 and W3\_FG39. W3\_FG65 is located at RA$ = 14^{\rm h}\ 10^{\rm m}\ 17.872^{\rm s}$, DEC$ = +57^\circ \ 41^\prime\ 39.815^{\prime\prime}$ (J2000), and its redshift is $z$=0.2514.
W3\_FG39 is located at RA $ = 14^{\rm h}\ 37^{\rm m}\ 36.402^{\rm s}$, DEC $ = +56^\circ \ 32^\prime\ 57.537^{\prime\prime}$ (J2000), and its redshift is $z=0.1492$. They are candidate FGs
in the sense that for these two groups no galaxy within the R$_{500}$ radius and within 2 magnitudes of that of the central galaxy is at the group redshift. However, X-ray data would be necessary to measure their
X-ray luminosity and confirm that they are bona fide FGs.
As is listed in Table~\ref{tab:spectro}, eight of our candidate FGs are in fact not FGs in the optical, and for the last three candidates there are missing redshifts, so our data do not allow us to come to any conclusions about their nature.

\subsection{The case of the W3\_FG39 BGG}
With the goal of adding FGs in Fig.~\ref{fig:analyse} (see below), we considered the two remaining FG candidates. The BGG of these groups being by definition the only bright galaxy of the systems, we were unable to
search for SbS within the group by lack of redshifts
and estimate their masses with the SG algorithm. However, the mass of W3\_FG39 was estimated 
to be $4.8 \times 10^{13}$~M$_{\odot}$ by \cite{Adami+20} (the mass of W3\_FG65 was not well constrained). 

The S/N of the W3\_FG39 BGG spectrum was the only one  high enough for FIREFLY to be applied to model its spectral energy distribution. As is expected for a FG central galaxy, FIREFLY basically requires a
single burst (91$\%$ of the produced stellar mass) 4~Gyrs aBB. Other bursts would only account for 9$\%$
of the mass, and would have occurred very early in the group history. We can note that the metallicity is the highest of the whole
studied sample: 2.1~Z$_{\odot}$.
These results (only a few old bursts, at high metallicity) were expected for a FG in a model in which FGs are formed by galaxy mergers in an isolated location.

\section{Discussion and conclusions}
\label{sec:discu}
In this work, we have used spectroscopic data to investigate the building behaviour of four galaxy groups at different stages of their evolution plus a candidate FG, investigating their internal SbS, which were detected using the Serna--Gerbal dendrogram method, and putting them in perspective with results from spectral energy distribution analyses of the member galaxies. We discuss our main results below. 

We discovered that the XCLASS~1330 galaxy system presents a peculiar distribution of SbS. 
 We detected a substructure (11) that is located outside the XMM-Newton X-ray emission region and has a short crossing time.
The XMM data do not cover this region of the sky, so we cannot determine whether there is an extension of the detected X-ray emission near substructure 11. However, with future X-ray surveys (e.g. by eROSITA) we shall be able to verify if substructure 11 exhibits a separate X-ray emission. This could indicate that it is an external galaxy structure falling into the central region of the XCLASS~1330 system, eventually becoming part of the future group. The other two SbS coincide with the X-ray emission and have a short crossing time, along with a high merging rate. Therefore, this system serves as an excellent laboratory for studying the build-up phase of galaxy groups.

Several works \citep[e.g.][]{2010MNRAS.405.1873D,2022ApJ...936...59D} have discussed the need to review the classification criteria of FGs. In this work, we demonstrate an issue with the magnitude gap definition using NGC 4104 as an example. It can be classified as a FG if the extended stellar envelope of the BGG is considered \citep[see our previous work by][]{LimaNeto+20}. However, when only the main component of the galaxy is taken into account, the magnitude gap criterium fails. This supports the discussion of a classification system redefinition. 

We detected only three SbS in NGC 4104. Given its relatively poor environment and its moderate distance from any cosmic node, it is unlikely that new galaxies will be accreted into the system. Therefore, despite the challenges in its fossil classification, there is no doubt that this group is in a more advanced stage of its evolution. Furthermore, its high metallicity and short crossing time suggest that intense past merging activity has already occurred, leading to significant metal diffusion throughout the system.

MCG+00-27-023 is a massive non-FG characterized by a considerable number of SbS and a history of regular infalling events. We predicted multiple burst episodes in the past within the detected SbS, some of which coincide with the merger event with the MCG+00-27-023 main structure. Furthermore, the current brightest and second brightest galaxies of the MCG+00-27-023 group do not satisfy the optical criterion ($\Delta m_{12} \geq 2.0$ mag) for fossil system classification, regardless of whether the extended envelopes are included or not. Given the predicted long crossing time in the core, it is unlikely to become a FG in the near future. However, we plan to investigate the optical criteria ($\Delta m_{12}$ or $\Delta m_{14}$) in future studies, taking into account the extended envelope of the galaxies using deep observations processed with a data reduction pipeline specially designed to preserve low surface brigthness structures, to determine whether the possibility of MCG+00-27-023 being a FG candidate should indeed be discarded (as is discussed in Appendix \ref{ap:decomp}).

The NGC 4065 group is a non-fossil system located in a dense environment near a cosmic node. We detected many SbS in this object (see Table \ref{tab:SGtable}), as is expected for a denser region, where the constant accretion of new members maintains the group dynamically active over long timescales. In particular, we identified two independent structures (112 and 1112), corresponding to small groups where the latest bursts of star formation, as was predicted by FIREFLY, occurred 13 Gyr aBB, suggesting that they recently merged with the NGC 4065 group. The substructure in the inner core (1111) does not present bursts in their recent star formation history. However, the crossing time estimated using the SG method is $\leq 0.1 \times 10^{9}$ years, suggesting that galaxy mergers should be highly frequent.

From our sample of FG candidates, we were able to confirm spectroscopically and photometrically the fossil nature of two candidates, including W3\_FG39. We found that the stellar mass of W3\_FG39 was assembled very early in its evolutionary history, with 91\% of the stellar mass produced only 4 Gyrs aBB. This suggests that the group experienced intense activity in the past and, as was expected for a fossil system, it now shows a decline in both dynamical and star formation activity. Therefore, investigating its large-scale environment could provide an insight into how this group evolved so quickly. Additionally, we expect to investigate the X-ray counterpart of W3\_FG39 in future studies.

\begin{figure}[ht]
\begin{center}
\includegraphics[width=0.4\textwidth]{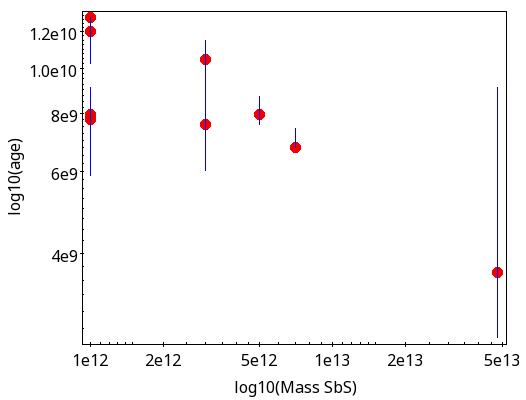}
\includegraphics[width=0.4\textwidth]{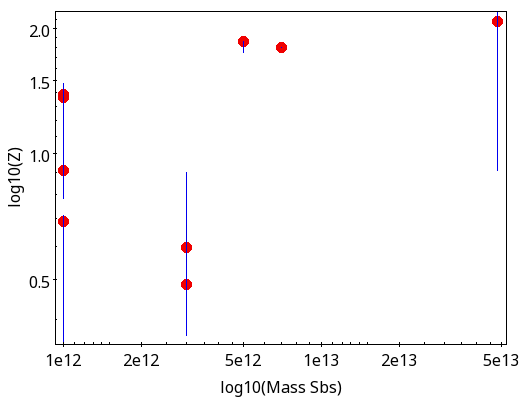}
\end{center}
\caption{Relations between the main parameters from Tables \ref{tab:SGtable} and \ref{tab:FIREFLY}. \textsf{Top:} Time elapsed aBB before the stellar population creation (in years: time aBB of the mean burst) versus SbS mass (in solar units). \textsf{Bottom:} 
Metallicity (in solar units) of the stellar population versus 
SbS mass. 
}
\label{fig:analyse}
\end{figure}

Correlating results from Tables \ref{tab:SGtable} and \ref{tab:FIREFLY}, we generated Fig.~\ref{fig:analyse} to investigate the relation between properties of the detected SbS. Uncertainties on the values plotted on these figures are quite  
large, but some tendencies appear, suggesting a continuous variation of the considered group properties.
First, the lower the mass of the SbS, the more recent the stellar population (meaning the burst occurred a long time after Big Bang). This is shown in the upper panel of Fig.~\ref{fig:analyse}. Computing the Kendall $\tau$ value to characterize the correlation (taking into account error bars of Fig.~\ref{fig:analyse}), we find $-0.67$, with an associated probability of 0.05. This approximately corresponds to a 2$\sigma$ level significance. Second, the higher the mass of the SbS, the higher the metallicity.
This is shown in the lower panel of Fig.~\ref{fig:analyse}. Computing similarly the Kendall $\tau$ value, we find $-0.61$ with an associated probability of 0.12. This approximately corresponds to a 1.5$\sigma$ level significance. This level is not significant but the figure can be interpreted as a lack of high-mass SbS for low-metallicity systems (empty lower right corner).
We also note that the property differences of regular and FGs are often not very significant \cite[see e.g.][for the 1-3$\sigma$ level difference for the galaxy orbits in fossil and regular groups]{2021A&A...655A.103Z}.

These results are consistent with the generally adopted model of energy transfer during interactions of the galaxies with the group and cluster potential wells \citep[e.g.][]{2000ASPC..197..377M}. The crucial parameter
in this process is the mass ratio between the parent structure and the galaxy. A small ratio, close to 1, as we find in small groups of galaxies, facilitates the energy transfer, and therefore the galaxy evolution. If the ratio is large, as in massive clusters, the energy transfer is inefficient, and the galaxy evolution is slow. This process is well known as the pre-processing of galaxies \citep[e.g.][]{2024MNRAS.528..919P} in low-mass groups, prior to their infall into massive clusters. 

In the present paper, the lowest-mass SbS exhibit galaxies that underwent relatively recent bursts of star formation under the likely effect of the SbS potential well. This would be due to the fact that the mass ratio between the SbS and the individual galaxies is low.
Similarly, the relatively high-mass SbS are probably older and have therefore been evolving for a longer time than low-mass SbS. This gave more time to the galaxies to enrich in terms of metallicity. 

Fossil groups were initially hypothesized to be the ultimate stage of galaxy group evolution \citep{1993Natur.363...51P}.
However, the FGs that we observe today will likely experience more evolution stages before becoming a single giant elliptical galaxy, 
surrounded by X-ray emission with a luminosity similar to that observed in groups or clusters, as is predicted by the $\Lambda$CDM cosmological model. Before reaching the end of the merger tree, the current FGs may lose their fossil status, according to the current classification system, if a bright galaxy falls into the group. Therefore, the timescale for a group to become a `true' FG is likely related to the large-scale environment in which it resides, and, of course, dynamical friction. From a dynamical perspective, we observed that the number of SbS tends to start from a lower number during group formation, increase over time, and then decrease as the group evolves into a fossil system.

In the future, we expect to analyse the positions of FGs and near-FGs in the cosmic web. By combining simulations and observational data, we plan to investigate the correlation between their cosmic web position with the number of internal SbS and their collapse timescale. In addition, we shall examine their neighbourhood to identify unbound structures that could merge with the groups in the future, allowing us to predict when a group might become a true fossil, and to better understand the process of a fossil system coming into being in the Universe. 

\begin{acknowledgements}
K.P.R. acknowledges financial support of the Coordenação de Aperfeiçoamento de Pessoal de Nível Superior (CAPES), Grant No. 88887.637628/2021-00 and 88887.694541/2022-00.
F.D. acknowledges long-term support from CNES. G.B.L.N. thanks the financial support from CNPq (grant 314528/2023-7) and FAPESP (grant 2024/06400-5).
This work is based on data obtained with the MISTRAL spectro-imager at Observatoire de Haute Provence (CNRS), France.
This research has made use of the MISTRAL database, operated at 
CeSAM (LAM), Marseille, France.

This work is also based on observations made in the Observatorios de Canarias del IAC with the TNG operated on the island of La Palma by the Fundaci\'on Galileo Galilei – INAF, Fundaci\'on Canaria in the Observatorio del Roque de los Muchachos.

Based on observations collected at Centro Astronomico Hispano en Andalucia (CAHA) at Calar Alto, operated jointly by Instituto de Astrofisica de Andalucia (CSIC) and Junta de Andalucia.

Based on observations taken with the Nordic Optical Telescope on La Palma (Spain). 

Based on observations obtained with the Telescopio Nazionale Galileo on La Palma (Spain).

This research has made use of the NASA/IPAC Extragalactic Database, which is funded by the National Aeronautics and Space Administration and operated by the California Institute of Technology.

This project used data obtained with the Dark Energy Camera (DECam), which was constructed by the Dark Energy Survey (DES) collaboration. Funding for the DES Projects has been provided by the US Department of Energy, the U.S. National Science Foundation, the Ministry of Science and Education of Spain, the Science and Technology Facilities Council of the United Kingdom, the Higher Education Funding Council for England, the National Center for Supercomputing Applications at the University of Illinois at Urbana-Champaign, the Kavli Institute for Cosmological Physics at the University of Chicago, Center for Cosmology and Astro-Particle Physics at the Ohio State University, the Mitchell Institute for Fundamental Physics and Astronomy at Texas A\&M University, Financiadora de Estudos e Projetos, Fundação Carlos Chagas Filho de Amparo à Pesquisa do Estado do Rio de Janeiro, Conselho Nacional de Desenvolvimento Científico e Tecnológico and the Ministério da Ciência, Tecnologia e Inovação, the Deutsche Forschungsgemeinschaft and the Collaborating Institutions in the Dark Energy Survey.

The Collaborating Institutions are Argonne National Laboratory, the University of California at Santa Cruz, the University of Cambridge, Centro de Investigaciones Enérgeticas, Medioambientales y Tecnológicas–Madrid, the University of Chicago, University College London, the DES-Brazil Consortium, the University of Edinburgh, the Eidgenössische Technische Hochschule (ETH) Zürich, Fermi National Accelerator Laboratory, the University of Illinois at Urbana-Champaign, the Institut de Ciències de l’Espai (IEEC/CSIC), the Institut de Física d’Altes Energies, Lawrence Berkeley National Laboratory, the Ludwig-Maximilians Universität München and the associated Excellence Cluster Universe, the University of Michigan, NSF NOIRLab, the University of Nottingham, the Ohio State University, the OzDES Membership Consortium, the University of Pennsylvania, the University of Portsmouth, SLAC National Accelerator Laboratory, Stanford University, the University of Sussex, and Texas A\&M University.

Based on observations at NSF Cerro Tololo Inter-American Observatory, NSF NOIRLab (NOIRLab Prop. ID 2022A-741884; PI: K. Parra Ramos), which is managed by the Association of Universities for Research in Astronomy (AURA) under a cooperative agreement with the U.S. National Science Foundation.

IM acknowledges financial support from the Severo Ochoa grant CEX2021-001131-S funded by MCIN/AEI/10.13039/501100011033, and the Spanish MCIU grant PID2022-140871NB-C21.

 The authors thank the referee for his/her constructive comments.
\end{acknowledgements}

\bibliographystyle{aa}
\bibliography{aa54570-25.bib}

\begin{appendix}
\onecolumn

\section{New redshift measurements}

In order to increase our redshift coverage as compared to the literature, we present in this section the redshift measurements that we made with several spectrographs.  We specifically targeted galaxies for which no spectroscopic redshift was available in NED. However,
we also targeted eight galaxies with known redshifts (seven with OHP/MISTRAL and one with CAHA/CAFOS) in order to estimate our measurement statistical error. We have a statistical difference between our own measurements and the NED spectroscopic redshifts of 0.0003$\pm$0.0025, with a maximum difference of 0.0068. This represents a mean value of 90 km~s$^{-1}$, of the same order as the redshift uncertainty due to the instrument wavelength calibration for OHP/MISTRAL\footnote{See page 6 of http://www.obs-hp.fr/guide/mistral/Test\_report.pdf}.

The redshift quality flags are the same as in \cite{Adami+18}. In a few words, flag 2 redshifts have a confidence estimated to be larger than 95$\%$, flags 3 and 4 are highly secure redshifts with a confidence larger than 99$\%$, and flag 9 are redshifts based on a single clear feature, given the absence of other features. 

\subsection{Redshifts for the XCLASS~1330, MCG+00-27-023, NGC~4065, and NGC~4104 groups}

\begin{table*}[ht]
\centering
\centering
\setcounter{table}{0}
\caption{New OHP/MISTRAL redshifts for the XCLASS~1330, MCG+00-27-023, NGC~4065, and NGC~4104 groups.}
\label{tab:NewMISTRAL}
\begin{tabular}{lcccc}
\hline \hline
Group & RA (J2000.0) & DEC (J2000.0) & z~~ & flag \\ 
\hline
XCLASS~1330 & 212.061137 & 78.615464 & 0.1985 & 2 \\
XCLASS~1330 & 212.049719 & 78.617114 & 0.1956 & 2 \\
XCLASS~1330 & 212.049718 & 78.610846 & 0.1981 & 2 \\
XCLASS~1330 & 212.032172 & 78.616948 & 0.1967 & 2 \\
XCLASS~1330 & 211.815376 & 78.614604 & 0.1952 & 3 \\
XCLASS~1330 & 212.267532 & 78.616625 & 0.1972 & 3 \\
XCLASS~1330 & 211.832387 & 78.629072 & 0.1943 & 3 \\
XCLASS~1330 & 211.821675 & 78.620270 & 0.1940 & 2 \\
XCLASS~1330 & 211.858057 & 78.646624 & 0.2148 & 2 \\
XCLASS~1330 & 211.975614 & 78.622421 & 0.     & 3 \\
XCLASS~1330 & 211.935977 & 78.633623 & 0.1994 & 3 \\ 
XCLASS~1330 & 211.937033 & 78.606812 & 0.1948 & 2 \\
XCLASS~1330 & 212.282065 & 78.611776 & 0.     & 3 \\
XCLASS~1330 & 212.307811 & 78.602660 & 0.     & 3 \\
XCLASS~1330 & 212.244435 & 78.592588 & 0.1911 & 2 \\ 
XCLASS~1330 & 212.271295 & 78.583906 & 0.2407 & 2 \\
XCLASS~1330 & 212.004132 & 78.594595 & 0.1933 & 2 \\
XCLASS~1330 & 212.072272 & 78.630674 & 0.2130 & 2 \\
XCLASS~1330 & 212.081825 & 78.636587 & 0.2118 & 2 \\
XCLASS~1330 & 212.124000 & 78.905060 & 0.0894 & 3 \\
XCLASS~1330 & 211.471540 & 78.386470 & 0.1998 & 2 \\
XCLASS~1330 & 211.810740 & 78.622177 & 0.1964 & 2 \\
XCLASS~1330 & 211.797840 & 78.630007 & 0.1963 & 3 \\
XCLASS~1330 & 211.874674 & 78.644178 & 0.1943 & 2 \\
XCLASS~1330 & 211.973541 & 78.569772 & 0.     & 3 \\
XCLASS~1330 & 211.976686 & 78.571950 & 0.     & 3 \\
XCLASS~1330 & 211.980440 & 78.604956 & 0.2655 & 2 \\
XCLASS~1330 & 212.265331 & 78.647055 & 0.1981 & 2 \\
\hline
MCG+00-27-023 & 157.479215 & -3.235516  & 0.0370 & 2\\            
MCG+00-27-023 &  157.565543&  -3.198520&  0.0369 & 2\\        
MCG+00-27-023 &  157.560890& -3.202506 & 0.0282 & 2\\               
MCG+00-27-023 & 157.523776 & -3.181808 & 0.0377 & 2\\               
MCG+00-27-023 &  157.408147 &  -3.189697 & 0.0951 & 2\\        
MCG+00-27-023 & 157.391370 &   -3.182189& 0.0377 & 2\\             
MCG+00-27-023 & 157.573694 &  -3.110431& 0.1389 & 2\\               
MCG+00-27-023 &   157.693477&  -3.168351& 0.0384 & 2\\             
MCG+00-27-023 & 157.677289 &   -3.164767& 0.0343 & 2\\            
MCG+00-27-023 &   157.527523&   -3.175079& 0.0356 & 2\\           
MCG+00-27-023 & 157.573475 & -3.103902 &  0.0376 & 2\\             
\hline
NGC~4065 & 181.450070 & 20.566786 & 0.0995 & 3 \\               
NGC~4065 & 181.407237 & 20.579238 & 0.0721 & 9 \\               
NGC~4065 & 181.445970 & 20.491991 & 0.1024 & 2 \\               
NGC~4065 & 181.448354 & 20.495691 & 0.0434 & 4 \\               
NGC~4065 & 181.482958 & 20.525536 & 0.0585 & 2 \\               
NGC~4065 & 181.488558 & 20.530269 & 0.0726 & 2 \\
\hline
NGC~4104 & 181.65422 & 28.186514 & 0.0283 & 9 \\ 
NGC~4104 & 181.91617 & 27.883472 & 0.0354 & 2 \\ 
\hline              
\hline
\end{tabular}
\tablefoot{
  The columns are: group name, J2000 right ascension and declination (in degrees), spectroscopic redshift,
  and redshift quality flag. Flag 2 redshifts have a confidence estimated to be larger than 95$\%$, flags 3
  and 4 are highly secure redshifts with a confidence larger than 99$\%$, and flag 9 are redshifts based on a
  single clear feature, given the absence of other features.
}
\end{table*}

\subsection{Redshifts for fossil group candidate galaxies of the CFHTLS}

\begin{table*}
\setcounter{table}{1}
\caption{Redshift measurements for FG candidate galaxies of the CFHTLS.}
\label{tab:spectro}
\begin{tabular}{rllllll}
\hline\hline
\multicolumn{7}{l}{Possible FGs} \\
\hline
Group & RA (J2000.0) & DEC (J2000.0) & r & z & flag & tel./instr. \\
\hline
W3\_FG65 & 212.574467 & 57.694393 & & 0.2514 & &  \\
1 & 212.54344 & 57.697403 & 18.491 & 0.2514 & 9 & NOT/ALFOSC \\
2 & 212.59952 & 57.671913 & 18.702 & 0.1041 & 4 & NOT/ALFOSC \\
3 & 212.62862 & 57.69387  & 18.992 & 0.3223 & 2 & CAHA/CAFOS \\
\hline
W3\_FG39 & 219.401673 & 56.549316 & & 0.1492 & &  \\
1 & 219.38786 & 56.56704  & 17.764 & 0.1492 & 4 & NOT/ALFOSC \\
2 & 219.41443 & 56.499195 & 18.474 & 0.2390 & 3 & NOT/ALFOSC \\
3 & 219.48647 & 56.562    & 18.658 & 0.2128 & 2 & OHP/MISTRAL \\
4 & 219.45607 & 56.586742 & 18.763 & 0.2245 & 2 & NOT/ALFOSC \\
5 & 219.35387 & 56.56855  & 18.815 & 0.2860 & 2 & NOT/ALFOSC \\
6 & 219.36154 & 56 52737  & 19.478 & 0.2713 & 2 & NOT/ALFOSC \\
\hline
\hline
\multicolumn{7}{l}{Non-FGs} \\
\hline
W1\_FG29 & 32.199306 & -6.595975 & & 0.1326 & &  \\
1 & 32.194866 & -6.5994077 & 17.135 & 0.1326 & 3 & OHP/MISTRAL \\
2 & 32.239437 & -6.5513453 & 18.223 & 0.1368 & 2 & CAHA/CAFOS \\
3 & 32.203796 & -6.5575233 & 18.495 & 0.1380 & 2 & OHP/MISTRAL \\
\hline
W2\_FG89 & 132.546320 & -3.923097 & & 0.3746 & & \\
1 & 132.56537 & -3.9137754 & 18.364 & 0.3746 & 3 & ATT \\
2 & 132.56467 & -3.9808667 & 18.975 & 0.2671 & 2 & CAHA/CAFOS \\
3 & 132.5028  & -3.9647741 & 19.036 & 0.3666 & 2 & CAHA/CAFOS \\
\hline
W2\_FG14 & 135.459925 & -4.325880 & & 0.1305 & & \\
1 & 135.45155 & -4.320919 & 17.424 & 0.1305 & 4 & OHP/MISTRAL \\
2 & 135.46371 & -4.357880 & 18.426 & 0.1373 & 2 & OHP/MISTRAL \\
3 & 135.49352 & -4.353849 & 18.591 & 0.1084 & 2 & CAHA/CAFOS \\
4 & 135.4871  & -4.342762 & 18.671 & 0.2120 & 9 & CAHA/CAFOS \\
\hline
W4\_FG47 & 335.335243 & 0.981269& &  0.2460 & & \\
1 & 335.36130 & 1.0062833 & 18.658 & 0.2472 & 2 & OHP/MISTRAL \\
2 & 335.36084 & 1.0062163 & 18.784 & 0.2461 & 2 & OHP/MISTRAL \\
\hline
W1\_FG122 & 36.759826 & -9.267997 & & 0.2225 & & \\
1  & 36.797894 & -9.252131 & 18.158 & 0.2225 & 3 & TNG/DOLORES \\
2  & 36.770134 & -9.258647 & 18.251 & 0.2609 & 9 & TNG/DOLORES \\
3  & 36.714222 & -9.266275 & 19.171 &        &  &    \\
4  & 36.787556 & -9.239616 & 19.389 & 0.2225 & 3 & TNG/DOLORES \\
5  & 36.761898 & -9.219572 & 19.537 & 0.3    & 3 & TNG/DOLORES \\
6  & 36.742523 & -9.22836  & 19.77  & 0.3167 & 2 & TNG/DOLORES \\
O9 & 36.737819 & -9.300233 &        & 0.265  & 2 & TNG/DOLORES \\
O11& 36.790792 & -9.230194 &        & 0.2859 & 2 & TNG/DOLORES \\
O6 & 36.745838 & -9.240942 &        & 0.2998 & 2 & TNG/DOLORES \\
O8 & 36.727159 & -9.289660 &        & 0.2444 & 3 & TNG/DOLORES \\
\hline
W1\_FG172 & 36.818780 & -4.550156 & & 0.215 & & \\
1 & 36.81482 & -4.5337305 & 17.842 & 0.215 & 3 & CAHA/CAFOS \\
2 & 36.78514 & -4.5476103 & 19.884 & 0.225 & 9 & CAHA/CAFOS \\
\hline
W1\_FG200 & 38.687428 & -5.641739 & & 0.1475 & & \\
1 & 38.703262 & -5.6659985 & 17.617 & 0.1475 & 3 & CAHA/CAFOS \\
2 & 38.68607  & -5.683833  & 19.404 &        &   &             \\
3 & 38.659294 & -5.6446605 & 19.595 &        &   &             \\
4 & 38.69901  & -5.6659775 & 19.796 & 0.1166 & 2 & CAHA/CAFOS \\
5 & 38.71499  & -5.680361  & 19.971 & 0.205  & 2 & TNG/DOLORES \\
6 & 38.69128  & -5.6154056 & 20.006 &        &   &               \\
\hline\hline
\end{tabular}
\tablefoot{
For each group, the first line gives the group name, approximate co-ordinates and BGG redshift. The following lines give the list of objects within 0.5 Virial radius (based on masses computed in \cite{Adami+20}) and in the two--magnitude range following the magnitude of the putative BGG. For each galaxy we give the co-ordinates, r band magnitude, spectroscopic redshift, quality flag (see Tab.~\ref{tab:NewMISTRAL}), and telescope and spectrograph where this redshift was obtained.  
}
\end{table*}

\begin{table*}
\setcounter{table}{1}
\caption{Continued.}
\begin{tabular}{rllllll}
\hline\hline
\multicolumn{7}{l}{Unconclusive}\\
\hline
Group & RA & DEC & r & z & flag & tel./instr. \\
\hline
W2\_FG95 & 133.388577 & -4.054261 & & 0.2638 & & \\
1   &  133.39284 & -4.0752935 &  18.690 & 0.220  & 3  & TNG/DOLORES \\
2   &  133.39392 & -4.0489383 &  18.804 & 0.363  & 3  & TNG/DOLORES \\
3   &  133.36308 & -4.0555100 &  19.006 & 0.482  & 9  & TNG/DOLORES \\
O6b &  133.36318 & -4.505667  &         & 0.0884 & 3  &  TNG/DOLORES \\
\hline
W4\_FG256 & 333.038432 & -0.269728 & & 0.426 & & \\
1 & 333.05182 & -0.2456301 & 19.023 & 0.426 & 9 & TNG/DOLORES \\
2 & 333.0909  & -0.2747156 & 20.609 & 0.479 & 9  & TNG/DOLORES\\
3 & 332.9991  & -0.2935538 & 20.715 &       &   \\
4 & 333.084   & -0.2822262 & 20.721 &       &   \\
5 & 332.99283 & -0.275597  & 20.777 & 0.525 & 2  & TNG/DOLORES\\
\hline
W4\_FG105 & 334.641805 & -0.756017 & & 0.243 & &  \\
1 & 334.6631  & -0.7753664 & 19.456 & 0.243 &  3 & TNG/DOLORES  \\
2 & 334.64697 & -0.7517933 & 19.487 & 0.329 &  3 & TNG/DOLORES \\ 
3 & 334.63086 & -0.7540495 & 19.702 & 0.1555&  1 & TNG/DOLORES  \\
4 & 334.68265 & -0.7457932 & 19.753 &       &     \\
5 & 334.67218 & -0.7620385 & 19.943 & 0.331 &  2 & TNG/DOLORES  \\
6 & 334.6024  & -0.7473659 & 19.991 & 0.499 &  1 & TNG/DOLORES  \\
O2& 334.613527& -0.724196  &        & 0.276 &  4  & TNG/DOLORES \\
\hline
W1\_FG153 & & & & 0.2770 & & \\
1 & 37.566944 & -4.480549  & 17.917 & 0.2770 & 3 & OHP/MISTRAL \\
2 & 37.542206 & -4.4316688 & 19.697 &        &   &             \\
\hline
\hline
\end{tabular}
\end{table*}


\section{Stellar populations derived with FIREFLY}

In order to estimate a typical uncertainty on the fits of FIREFLY due to different models, we used two different stellar libraries to fit our summed spectra: MILES and STELIB.
Table~\ref{tab:FIREFLY} gives the results of the two models.

We first computed the width of the error bars for the metallicity Z and the extinction E(B-V) as compared to the Z and E(B-V) mean values. The FIREFLY internal error bars on Z and E(B-V) represent 22$\%$ and
24$\%$ of the mean values. 

We then computed the value of the difference between the two stellar libraries (MILES and STELIB) that we considered for Z, E(B-V), and the main age, as compared to the mean values coming from the MILES model. The difference between the values of E(B-V) between the models obtained with MILES and STELIB is equal to (23$\pm$18$)\%$ of E(B-V) when considering the MILES library.  For the main age and Z, it is (27$\pm$48$)\%$ and (1$\pm$18$)\%$ (if we exclude from the sample the three lowest metallicities).

As a conclusion, we can say that the typical uncertainties of FIREFLY
on E(B-V), age, and Z, can be considered of the order of 0.1, 3~Gyrs and 0.2 Z$_\odot$.

\begin{landscape}    
\begin{table}[ht]
\tiny
\caption{Stellar population synthesis results obtained with FIREFLY.
}
\label{tab:FIREFLY}
\begin{tabular}{lccccccccccccccccccc}
\hline \hline
Name & E(B-V) & age & age$_{up}$ & age$_{low}$ & Z & Z$_{up}$ &  Z$_{low}$ & 
ages$_{0}$ & w$_{0}$ & ages$_{1}$ & w$_{!}$ & 
ages$_{2}$ & w$_{2}$ & ages$_{3}$ & w$_{3}$ & 
ages$_{4}$ & w$_{4}$ & ages$_{5}$ & w$_{5}$ \\
\hline
\hline
MCG\_Main             & 0.226 &    0.83 &    0.92 &    0.79 &  -0.173 &  -0.119 &  -0.314 &   1.000 &   0.485 &   0.602 &   0.242 &   0.301 &   0.227 &   1.041 &   0.045 & -99.990 & -99.990 & -99.990 & -99.990 \\
MCG\_12               & 0.445 &    0.88 &    0.96 &    0.78 &  -0.314 &  -0.173 &  -0.436 &   0.845 &   0.273 &   0.954 &   0.273 &   1.079 &   0.181 &   0.176 &   0.181 & 0.778 & 0.045 & 1.079 & 0.045 \\
MCG\_111112           & 0.000 &    1.08 &    1.09 &    1.01 &  -0.041 &   0.090 &  -0.106 &   1.114 &   0.371 &   1.000 &   0.371 &   1.114 &   0.248 &   1.114 &   0.010 & -99.990 & -99.990 & -99.990 & -99.990 \\
MCG\_1111111          & 0.219 &    1.02 &    1.06 &    0.95 &  -0.224 &  -0.045 &  -0.257 &   1.000 &   0.500 &   1.000 &   0.250 &   1.079 &   0.250 & -99.990 & -99.990 & -99.990 & -99.990 & -99.990 & -99.990 \\
\hline
NGC~4065\_Main         & 0.106 &    0.77 &    0.85 &    0.71 &   0.110 &   0.180 &  -0.034 &   0.954 &   0.300 &   0.602 &   0.300 &   0.954 &   0.200 &   0.000 &   0.200 &  -2.046 &   0.001 & -99.990 & -99.990 \\
NGC~4065\_112          & 0.166 &    0.90 &    0.96 &    0.77 &  -0.163 &  -0.149 &  -0.633 &   1.114 &   0.250 &   1.079 &   0.250 &  -0.222 &   0.250 &   0.778 &   0.250 & -99.990 & -99.990 & -99.990 & -99.990 \\
NGC~4065\_1112         & 0.083 &    0.89 &    0.93 &    0.83 &   0.142 &   0.169 &   0.087 &   0.602 &   0.455 &   1.114 &   0.227 &   1.079 &   0.227 &   0.301 &   0.091 & -99.990 & -99.990 & -99.990 & -99.990 \\
\hline
NGC~4104\_Main         & 0.045 &    0.93 &    0.97 &    0.88 &   0.254 &   0.254 &   0.201 &   1.000 &   0.500 &   1.079 &   0.250 &   0.301 &   0.250 & -99.990 & -99.990 & -99.990 & -99.990 & -99.990 & -99.990 \\
NGC~4104\_11          & 0.053 &    0.83 &    0.87 &    0.83 &   0.254 &   0.254 &   0.254 &   1.079 &   0.250 &   0.699 &   0.250 &   0.176 &   0.167 &   0.602 &   0.167 &   1.000 &   0.167 & -99.990 & -99.990 \\
NGC~4104\_err11      & 0.045 &    0.85 &    0.92 &    0.83 &   0.254 &   0.256 &   0.224 &   1.079 &   0.250 &   0.176 &   0.250 &   0.602 &   0.250 &   1.041 &   0.250 &  -2.046 &   0.001 & -99.990 & -99.990 \\
NGC~4104\_12          & 0.068 &    0.90 &    0.94 &    0.88 &   0.270 &   0.270 &   0.243 &   1.041 &   0.303 &   0.301 &   0.303 &   1.079 &   0.303 &   0.602 &   0.091 & -99.990 & -99.990 & -99.990 & -99.990 \\
NGC~4104\_2            & 0.008 &    1.11 &    1.11 &    1.08 &   0.135 &   0.135 &   0.128 &   1.114 &   0.495 &   1.114 &   0.495 &  -0.523 &   0.010 & -99.990 & -99.990 & -99.990 & -99.990 & -99.990 & -99.990 \\
\hline
W3\_FG39\_BCG          & 0.068 &    0.56 &    0.96 &    0.42 &   0.318 &   0.349 &  -0.041 &   0.602 &   0.909 &  -1.000 &   0.091 & -99.990 & -99.990 & -99.990 & -99.990 & -99.990 & -99.990 & -99.990 & -99.990 \\
\hline
\hline
MCG\_Main             & 0.173 &    0.49 &    0.59 &    0.44 &  -0.317 &  -0.317 &  -0.327 &   0.301 &   0.660 &   0.699 &   0.330 &   1.041 &   0.010 & -99.990 & -99.990 & -99.990 & -99.990 & -99.990 & -99.990 \\
MCG\_12               & 0.392 &   0.070 &   0.281 &   -0.018 &  0.156 &  0.156 &  0.113 &  -0.046 &   0.455 &  0.176 &   0.455 &  -0.097 &   0.091 & -99.990 & -99.990 & -99.990 & -99.990 & -99.990 & -99.990 \\
MCG\_111112           & 0.000 &    0.35 &    0.72 &    0.22 &  -0.151 &  -0.059 &  -0.327 &   0.301 &   0.455 &   0.176 &   0.455 &  -0.046 &   0.045 &   1.114 &   0.045 & -99.990 & -99.990 & -99.990 & -99.990 \\
MCG\_1111111          & 0.219 &    1.01 &    1.05 &    1.01 &  -0.201 &  -0.201 &  -0.327 &   1.114 &   0.682 &   0.602 &   0.227 &   0.778 &   0.091 & -99.990 & -99.990 & -99.990 & -99.990 & -99.990 & -99.990 \\
\hline
NGC~4065\_Main         & 0.113 &    0.32 &    0.58 &    0.32 &   0.047 &   0.210 &   0.047 &   0.301 &   0.909 &   0.477 &   0.091 & -99.990 & -99.990 & -99.990 & -99.990 & -99.990 & -99.990 & -99.990 & -99.990 \\
NGC~4065\_112          & 0.121 &    0.42 &    0.46 &    0.38 &   0.089 &   0.111 &   0.076 &   0.477 &   0.364 &   0.477 &   0.364 &  -0.301 &   0.182 &   0.602 &   0.030 &   0.845 &   0.030 &  -0.097 &   0.030 \\
NGC~4065\_1112         & 0.045 &    0.99 &    1.03 &    0.99 &   0.272 &   0.276 &   0.263 &   1.041 &   0.792 &   0.699 &   0.198 &  -0.097 &   0.010 & -99.990 & -99.990 & -99.990 & -99.990 & -99.990 & -99.990 \\
\hline
NGC~4104\_Main         & 0.000 &    1.08 &    1.10 &    1.08 &   0.251 &   0.252 &   0.251 &   1.114 &   0.495 &   1.079 &   0.248 &   1.041 &   0.248 &  -0.097 &   0.010 & -99.990 & -99.990 & -99.990 & -99.990 \\
NGC~4104\_11         & 0.030 &    0.98 &    0.98 &    0.97 &   0.241 &   0.241 &   0.233 &   1.114 &   0.500 &   0.845 &   0.167 &   0.301 &   0.167 &   1.000 &   0.167 & -99.990 & -99.990 & -99.990 & -99.990 \\
NGC~4104\_err\_11    & 0.030 &    1.05 &    1.05 &    1.04 &   0.215 &   0.232 &   0.215 &   1.114 &   0.660 &   0.903 &   0.330 &  -0.097 &   0.010 & -99.990 & -99.990 & -99.990 & -99.990 & -99.990 & -99.990 \\
NGC~4104\_12         & 0.075 &    0.97 &    0.97 &    0.95 &   0.347 &   0.349 &   0.347 &   0.778 &   0.495 &   1.114 &   0.495 &  -0.222 &   0.010 & -99.990 & -99.990 & -99.990 & -99.990 & -99.990 & -99.990 \\
NGC~4104\_2            & 0.008 &    1.07 &    1.08 &    1.00 &   0.156 &   0.254 &   0.130 &   1.041 &   0.455 &   1.114 &   0.455 &   0.954 &   0.091 & -99.990 & -99.990 & -99.990 & -99.990 & -99.990 & -99.990 \\
\hline
W3\_FG39\_BCG         & 0.151 &    0.35 &    0.74 &    0.18 &  -0.133 &   0.047 &  -0.327 &   0.477 &   0.500 &   0.176 &   0.500 & -99.990 & -99.990 & -99.990 & -99.990 & -99.990 & -99.990 & -99.990 & -99.990 \\
\hline
\hline
\end{tabular}
\label{tab:firefly}
\tablefoot{The first eight columns are:
Galaxy name, E(B-V), age (time elapsed aBB before the mean burst), maximum age, minimum age, metallicity, maximum metallicity, minimum metallicity. Ages are in units of log(age/Gyr) and metallicities are expressed as log(Z/H), with values of -99.990 for bursts that are not needed in the fit.
The next columns show the ages and weights (w) for the six single bursts. 
The top and bottom tables correspond to the results obtained with the MILES and STELIB libraries, respectively.
}
\end{table}

\end{landscape}

\section{MCG+00-27-023 group photometric decomposition}
\label{ap:decomp}
We used deep optical observations obtained with the Dark Energy Camera (DECam) at the Victor M. Blanco Telescope (Proposal ID: \#2022A-741884, PI: Kethelin Parra Ramos) in the $r$-band to perform a 2D surface brightness decomposition of the brightest (MCG+00-27-023) and second-brightest (LEDA 30995) galaxies in the MCG+00-27-023 group. This analysis aims to improve the photometric accuracy and to double-check the non-fossil status of this group. The data were reduced by the DECam Community Pipeline \citep{2014Valdes}, and the surface brightness limit value of the image is $\mu^{lim}_{r}(3\sigma, 10''\times10'')= 29.85$ mag arcsec$^{-2}$. 

The final image was masked and sky-subtracted (using a $3\sigma_{\mathrm{MAD}}$\footnote{Median absolute deviation (MAD)} median clipping of the fully masked image) before the decomposition. The masks were created using the software \texttt{NoiseChisel} and \texttt{Segment} \citep{2015AkhlaghiTakashi,2019arXiv190911230A} in the image in which the target galaxy had already been subtracted. The subtraction of the target galaxy was performed using the \texttt{Ellipse} function from the \textsc{photutils} package \citep{larry_bradley_2024_13989456}, in order to avoid its detection by \texttt{NoiseChisel}. 

\begin{table}[h]
    \centering
    \caption[]{Best-fit parameters for the two Sérsic components fitted with GALFIT.}
    \begin{tabular}{lcccccc}
     \hline \hline
      &  \multicolumn{6}{c}{Sérsic 1}  \\
        \cmidrule(lr){2-7} 
      Name & $m_{1}$ &$R_{\mathrm{eff},1}$ & $n_{1}$ & $b/a_{1}$ & PA$_{1}$ & C0$_{1}$ \\
    & \scriptsize[mag] &  \scriptsize[kpc] & & &\scriptsize[deg]&\\       
   (1) & (2) & (3) & (4) & (5) &(6) &(7) \\ 
   \hline \\
MCG+00-27-023&13.454 $\pm$ 0.001&5.34 $\pm$ 0.01&3.614 $\pm$ 0.003 &0.7242 $\pm$ 0.0001&38.79 $\pm$ 0.01&-0.088 $\pm$ 0.001\\ 
   \\
LEDA 30995&18.009 $\pm$ 0.003&0.314 $\pm$ 0.001&1.19 $\pm$ 0.02&0.407 $\pm$ 0.002&-89.9 $\pm$ 0.2&-\\
   
   \hline
      &  \multicolumn{6}{c}{Sérsic 2}  \\
        \cmidrule(lr){2-7} 
       & $m_{2}$  &$R_{\mathrm{eff},2}$ & $n_{2}$ & $b/a_{2}$ & PA$_{2}$ & C0$_{2}$ \\
    & \scriptsize[mag] &  \scriptsize[kpc] & & &\scriptsize[deg]& \\       
    &(8)&(9) &(10)&(11)&(12)&(13)\\ 
   \hline \\
MCG+00-27-023&13.425 $\pm$ 0.001&37.21 $\pm$ 0.03&1.223 $\pm$ 0.001& 0.5927 $\pm$ 0.0002&44.41 $\pm$ 0.02&-0.274 $\pm$ 0.001\\
   \\
LEDA 30995& 14.1968 $\pm$ 0.0004&7.55 $\pm$ 0.01 &3.856 $\pm$ 0.002 &0.8859 $\pm$ 0.0002 &34.9 $\pm$ 0.1&-\\
   
   \hline   
    \end{tabular}
    \label{tab:galfit_output}
\tablefoot{The columns (2) and (8) are the best-fitted magnitudes of the two Sérsic components from \textsc{galfit}. The columns (3) and (9) are the effective radius. The columns (4) and (10) are the Sérsic index. The columns (5) and (11) are the axis ratio. The columns (6) and (12) are the position angle. The parameter C$0$ in the columns (7) and (13) represents the diskyness ($<0$) or boxyness ($>0$) of the model.}\\    
\end{table}

\begin{table}[h!]
\addtocounter{table}{-1}
\caption{Continued.}                
\label{tab:galfit_output2}
\centering                       
\begin{tabular}{lccccc}      
\hline\hline              
Name & $m_{1}^{\mathrm{corr}}$&$m_{2}^{\mathrm{corr}}$ & $m_{T}^{corr}$ &  $\chi^{2}_{\mathrm{red}}$ \\ 
&[mag]&[mag] &[mag]& \\
(1) &(2)&(3) &(4)&(5)\\ 
\hline                     
MCG+00-27-023 & 13.340 $\pm$ 0.002&13.311 $\pm$ 0.002&12.573 $\pm$ 0.001  & 1.28\\  
LEDA 30995&17.884 $\pm$ 0.003&14.072 $\pm$ 0.001&14.040 $\pm$ 0.001&1.27  \\
\hline                                 
\end{tabular}
\tablefoot{The columns (2) and (3) are k- and dust reddening corrected magnitudes from the two Sérsic components magnitudes ($m_{1}$ and $m_{2}$) listed in de Table \ref{tab:galfit_output}. The column (4) is the integrated corrected magnitude of the galaxy, and the column (5) is the reduced $\chi^{2}$. }\\  
\end{table}

We used the \textsc{galfit} software \citep{2002Peng,2010Peng} for the multi-component fitting on the masked and sky-subtracted image. Table \ref{tab:galfit_output} shows the best-fit parameters for the two Sérsic components fitted to both the brightest and second-brightest galaxies. We applied the k-correction \citep{2010Chilingarian,2012Chilingarian} and dust reddening correction \citep{2011Schlafly} in the magnitudes shown in Table \ref{tab:galfit_output2}. The PSF was taken into account.

The magnitude gap between the brightest and second-brightest galaxy of the MCG+00-27-023 group is $\Delta m_{12}= 1.47$ mag. 
Therefore, it is not classified as a FG. The standard data reduction pipelines are not suitable for preserving low surface brightness light, often over-subtracting the sky background and introducing artefacts (e.g. dark halos around bright sources), as is the case with the DECam Community Pipeline.
As a result, a fraction of the extended stellar halo surrounding the galaxies was lost. However, our image is deep enough to preserve a portion of this diffuse halo. Therefore, we assumed the missing fraction of the extended halo would not significantly affect the fitted magnitudes. Nevertheless, this analysis should be revisited in the future using a pipeline designed for low surface brightness studies.

\section{NGC 4098 data analysis}
\label{ap:ngc4098}

Discovered by W. Herschel in 1785, the system known as VV 61 in the Vorontsov-Velyaminov catalogue of interacting galaxies has been classified by these authors as a E+E? pair in contact (PK), probably due to the roundish appearance of the galaxies at high surface brightness.  

It seems that the NGC name of the pair is confusing:  the main object to the north is referred to as NGC 4098-1 or NGC 4099-1 and the southern one as NGC 4098-2 or NGC 4099-2.  UGC 07091 is the unique name found in the Uppsala catalogue for both NGC~4098 and NGC~4099.
In this paper, we call NGC 4098 the northern one and NGC 4099 the southern one.  NGC 4098 is visually classified as Sab, and NGC 4099 has no known classification, but is similar to a roundish compact galaxy in the model displayed in the middle-top image. All the images show extended diffuse light around the pair of galaxies. The Hubble distance at the velocity of group G112 (7280 km s$^{-1}$) is $\sim$ 105 Mpc, leading to a pixel scale of $\sim$0.51 kpc arcsec$^{-1}$.

\begin{figure}[ht]
    \centering
        \includegraphics[width=\textwidth]{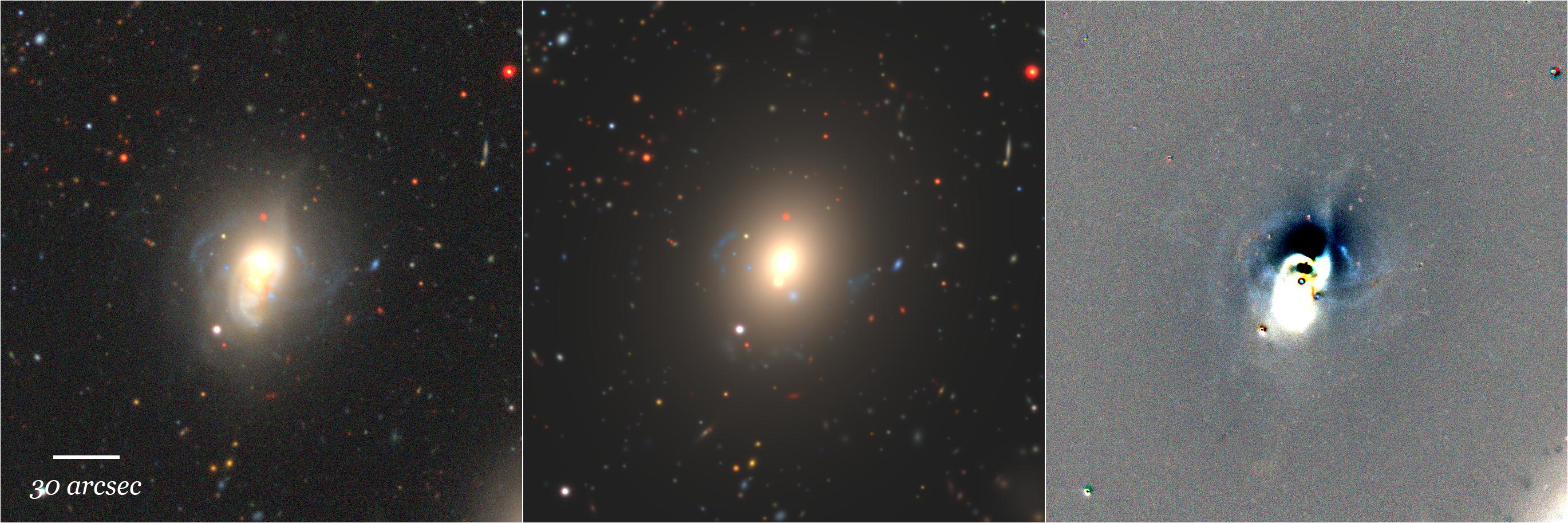} 
        \includegraphics[width=0.275\textwidth]{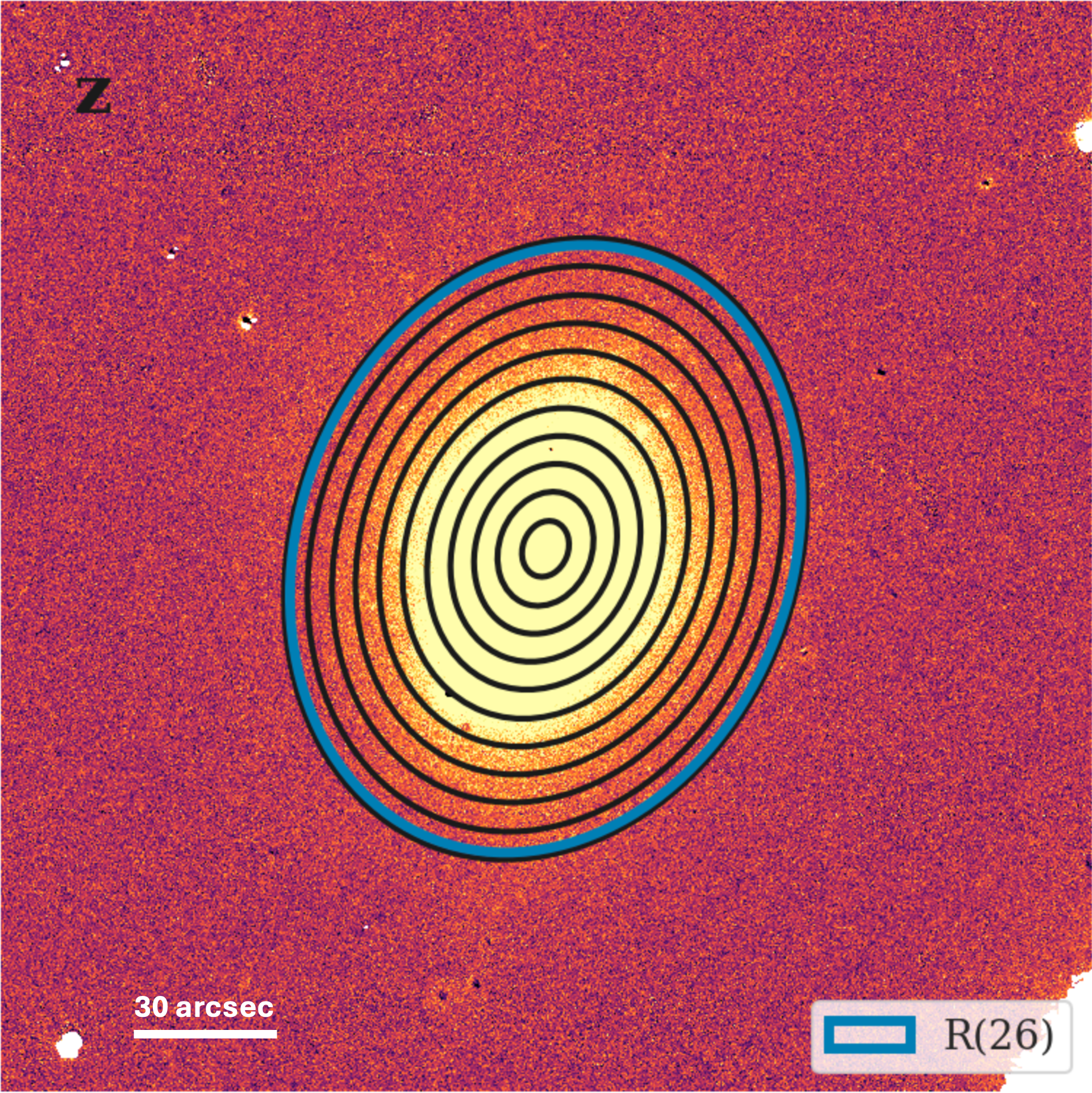} 
        \includegraphics[width=0.234\textwidth]{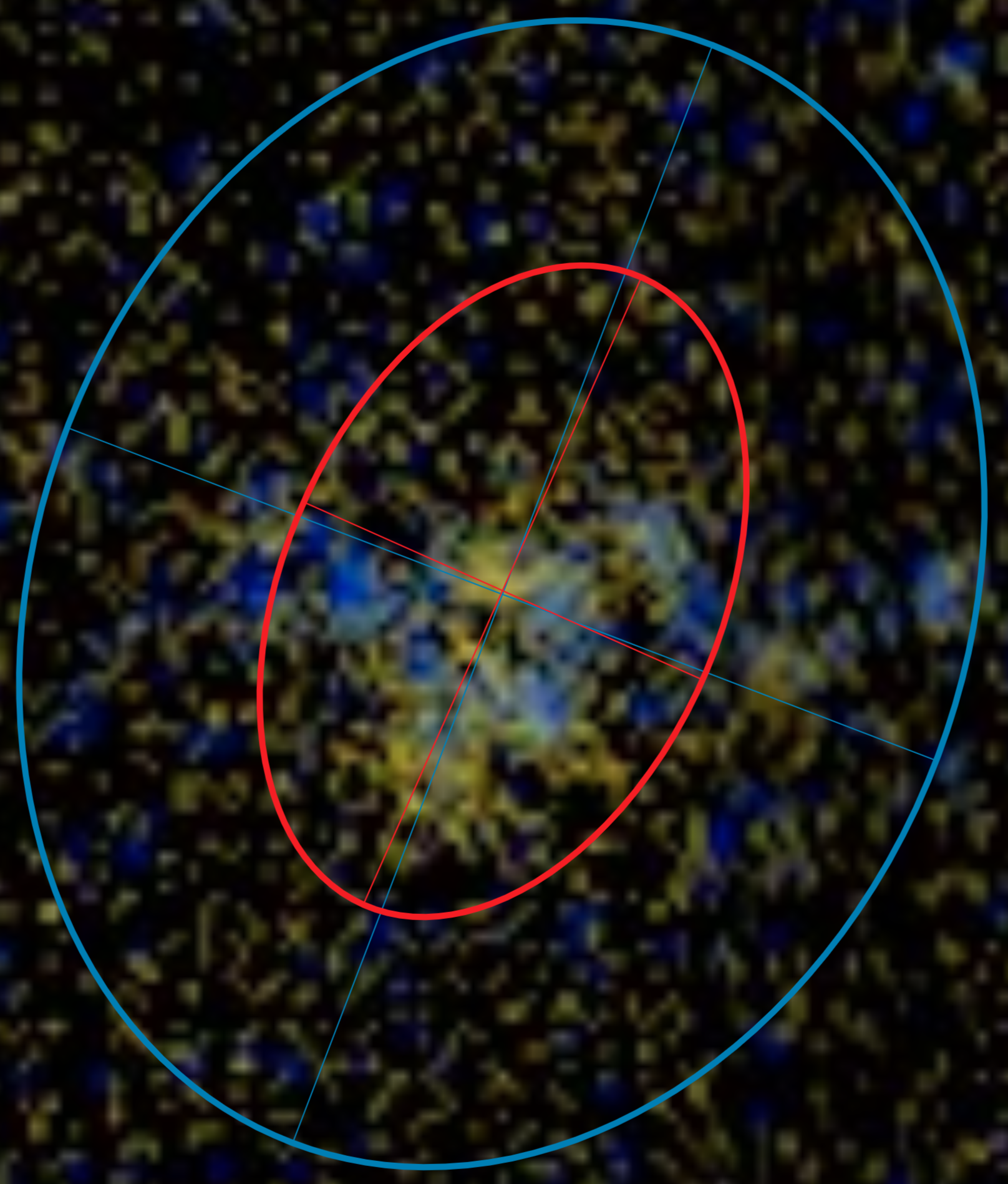} 
        \includegraphics[width=0.234\textwidth]{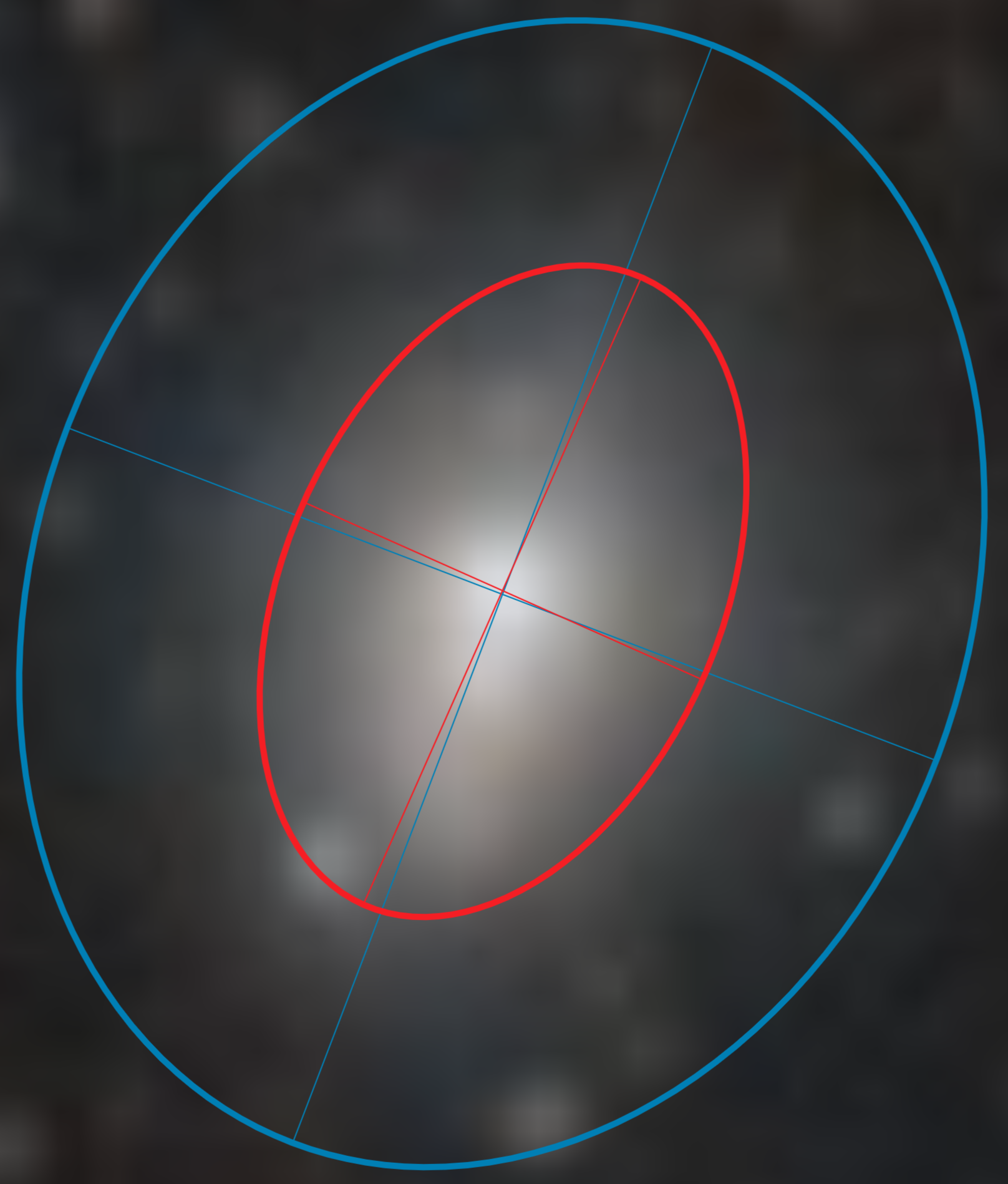} 
        \includegraphics[width=0.234\textwidth]{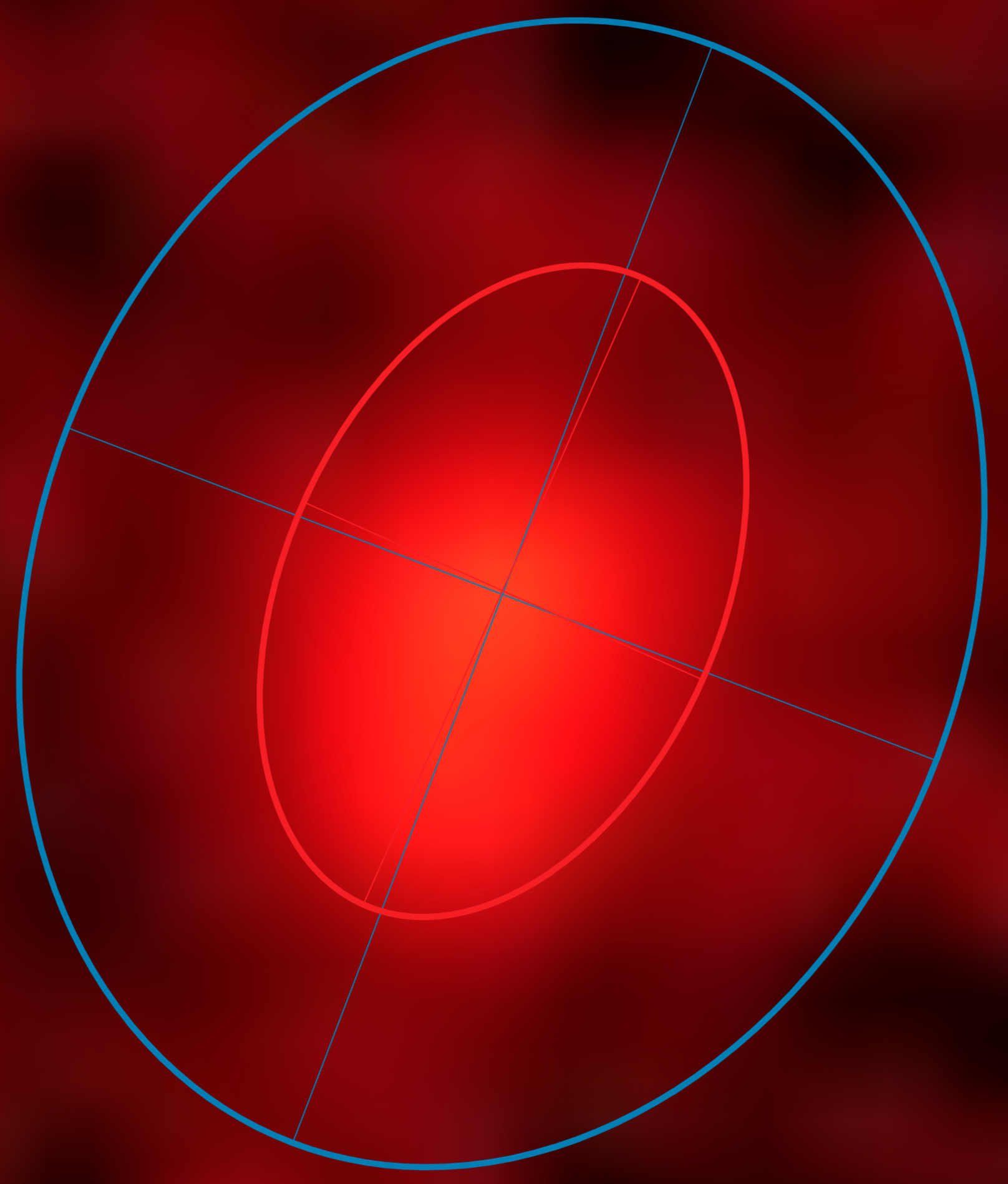} 
    \caption{
    Images of the galaxy NGC 4098 from the SGA legacy Survey (SGA ID 1177446).  
    Top line: g-, r-, z-band mosaic images.  From left to right : (1) data, (2) model and (3) residuals.  
    Bottom line, from left to right: (1) deep z-band DR10, (2) GALEX, (3) unWISE W1/W2 N07, and (4) WISE 12 micron dust map.  
    The blue ellipses come from the Siena Galaxy Atlas and model the surface brightness threshold radii at 26 mag arcsec$^{-2}$. The scales of the three last images (2), (3), and (4) are given by the blue ellipse, which is the same as for the first image (1).
    The red ellipses are extracted from the HyperLEDA SGA.
    }
    \label{fig:NGC4098-largegalaxy-grz-montage}
\end{figure}

Elongated structures surrounding the system have been adjusted by ellipses both in the R-band (red ellipses, Hyperleda) and in the z-band (blue ellipses), as displayed in the panel bottom row of Fig. \ref{fig:NGC4098-largegalaxy-grz-montage}. The position angles (PA) of the major axes of the ellipses are $\sim$156$^\circ$ and $\sim$159$^\circ$ respectively, and the inclinations with respect to the sky plane $\sim$40$^\circ$ and $\sim$52$^\circ$ respectively. The surface brightness profiles, in any band (g, r and z), follow almost a pure $r^{1/4}$ de~Vaucouleurs law, indicating that the elliptical shape of the whole system is an advanced relaxation phase.

As displayed in Fig. \ref{fig:ngc4098_VF}, the monochromatic flux map is a combination of data cubes obtained from different smoothings of 3, 5, and 7 pixels ($\sim$ 2.0, 3.4, and 4.7 arcsec respectively). In regions where the H$\alpha$ flux is most intense, the H$\alpha$ map comes from the least smoothed cube (3$\times$3 px$^2$). Below a certain flux threshold, the 5$\times$5 px$^2$ smoothed cube is used, and finally, below another lower threshold, the 7$\times$7 px$^2$ smoothed cube is used.  The continuum emission is the underlying emission below the H$\alpha$ line, within the passband of the interference filter (FWHM of 15~\AA). A Gaussian spatial smoothing of 9 pixels ($\sim$6 arcsec) was used to produce the continuum map, in order to reach the most external regions of the system, tracing an elongated structure.

We note three large blobs of H$\alpha$ emission, almost equidistant, aligned along an almost North-South axis (PA=350$^\circ$), the most intense being in the north and the least intense in the south. The northernmost blob corresponds, with a shift of about $\sim$1 kpc ($\sim$2~arcsec) towards the east, to the maximum emission of the continuum flux from the galaxy NGC 4098. The blob in the middle corresponds to the galaxy NGC 4099, and one can see in the deformed continuum isocontours at this location an underlying stellar component, as it was already the case on three panels of \ref{fig:NGC4098-largegalaxy-grz-montage}. The southernmost blob, on the contrary, does not present any visible stellar counterpart on the continuum map. The right panel of Fig. \ref{fig:ngc4098_VF} shows the velocity field corresponding to the monochromatic H$\alpha$ image, zoomed by a factor of 2, shown in the left panel.

A tiny line-of-sight velocity gradient is observed through the main body of the NGC 4098 galaxy, extending over $\sim$7.5 kpc ($\sim$15~arcsec), as measured from the central surface brightness profiles from the broadband imaging (Fig. \ref{fig:NGC4098-largegalaxy-grz-montage}) and from the H$\alpha$ map. The maximum velocity amplitude along the velocity field is 38 km s$^{-1}$, which leads to a gradient of $\sim$5 km s$^{-1}$ kpc$^{-1}$. This low value could be due to a low inclination.  Using the model describing the surface brightness distribution of the galaxy shown in the top-middle panel of Fig. \ref{fig:NGC4098-largegalaxy-grz-montage}, we measure an inclination of 57$\pm$1$^\circ$, which leads to a maximum amplitude for the rotation curve of V$_{max}(r=3.5\rm{ kpc})$=21$\pm$1 km s$^{-1}$ and a mass within a radius of 3.75 kpc of M(r$<$3.5kpc)$\sim$ 3.9 $\times 10^{8}$ M$_\odot$.  However, this result strongly depends on the very uncertain inclination of the galaxy. Following the Tully-Fisher relation (Tully \& Fisher 1977), a galaxy with a $R_{25}$ radius of 3.75 kpc can have a baryonic mass of M(r$<$3.5kpc)$\sim$9.5 $\times 10^{9}$ M$_\odot$ \citep[see Fig. 4 in][]{2011MNRAS.416.1936T}.
Thus, if NGC 4098 has an inclination of 9.6$^\circ$ instead of 57$^\circ$, then V$_{max}\sim$104 km s$^{-1}$ and the mass jumps by a factor of $\sim$24 !

A spatial extension is observed east of NGC 4099 at a radial velocity of 7600 km~s$^{-1}$, which is not in velocity continuity with the rest of the global velocity field, but the S/N is low. 
A smooth line-of-sight velocity continuity is observed between the north and south of the velocity field, suggesting that one faces a single kinematic structure extending from north to south on $\sim$22.3 kpc ($\sim$44 arcsec), with a velocity gradient of $\sim$295 km s$^{-1}$ (from  $\sim$6948 km s$^{-1}$ in the north on NGC 4098 to  $\sim$7243 km s$^{-1}$ to the south).  This velocity gradient can be produced by streaming motions during the interaction between galaxies. If this structure is a tidal tail, the large gradient indicates that it could not be in the sky plane but could develop a tridimensional pattern.  The observation of the tidal tail can be affected by projection effects, as suggested by the fact that the maximum rotation velocity of the tail is not observed at the end of the tail, where the velocity is $\sim$70~km~s$^{-1}$ lower, but $\sim$5~kpc ($\sim$10~arcsec) toward its end, as shown in Figs. 4, 5, and 6 of \cite{2004A&A...425..813B}.
Consequently, the tail could be longer than observed in projection. The difference between the peak velocity within the tail and the systemic velocity of NGC 4098 is $\sim$220 km s$^{-1}$ for a distance separating them of d$\sim$11kpc, requiring a time of $\sim$49 Myr to cover the distance.  This is a very rough estimate because one can only measure projected distances on the sky plane and velocities along the line-of-sight, and not 3D vectors.  Despite that, the actual length of the tail, $d$, measured in 3D is actually larger than its projection on the sky, while the real velocity $v$ is also larger that its line-of-sight component. These two effects combined mean that they compensate each other when calculating characteristic times (t=d/v), to some extent.

\begin{figure}[ht]
    \hspace*{-2cm}
    \includegraphics[width=1.2\textwidth]{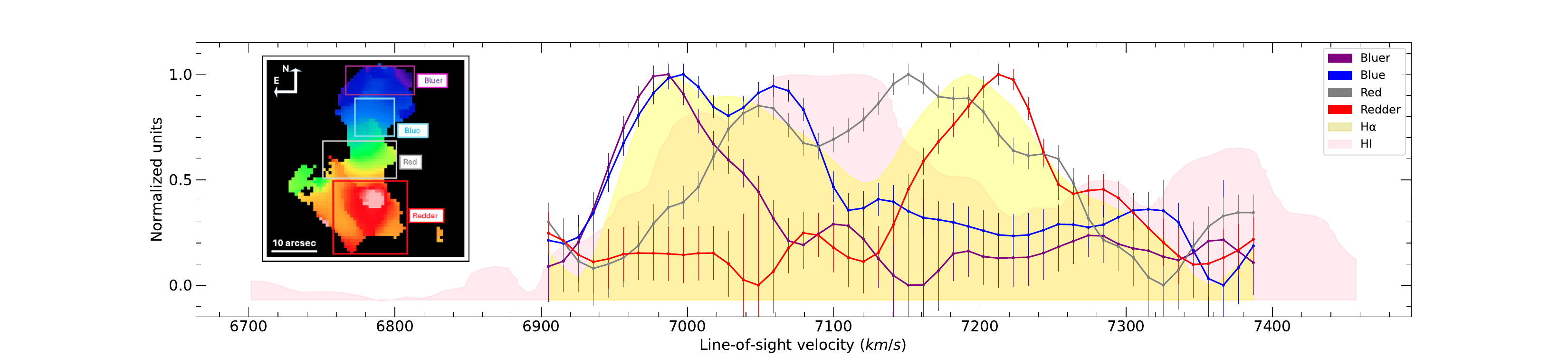} 
    \caption{%
        Velocity profiles integrated over different fields of view. The purple, blue, grey, and red profiles represent the H$\alpha$ emission lines from the less redshifted northern regions to the more redshifted ones in the south, called bluer, blue, red, and redder, respectively. They are integrated into the rectangular boxes of the same colour shown in the inset on the left of the figure. This velocity field, is used here as a reference, and is the one already displayed on the left panel of Fig. \ref{fig:ngc4098_VF}. The profile in yellow represents the H$\alpha$ line integrated over the entire object. Only profiles where a velocity could be calculated were used in the integrated profile. The profile in pink is the integrated HI from the  ALFALFA survey profile. All profile amplitudes have been normalized to unity. Error bars are Poissonian.
    }
    \label{fig:ngc4098_VelocityProfiles}
\end{figure}

Figure \ref{fig:ngc4098_VelocityProfiles} complements Fig.  \ref{fig:ngc4098_VF}; the first four curves (respectively named `bluer', `blue', `red' and ``redder'') still show this velocity gradient from north to south but illustrate better that the ‘‘blue‘‘ component clearly shows two components, the one found in the ‘bluer' component but also the one that we will find at the lowest velocities in the `red' component.  We also note that the most redshifted component was already guessed in the `bluer' component. The `red' component spreads in velocity, up to values almost equal to that found in the `redder' component, and shows a large velocity amplitude (FWHM $\sim$250 km s$^{-1}$ around the galaxy NGC 4099). We also note that the `redder' component shows, like the `bluer' component, an unresolved double profile. The yellow surface named H$\alpha$ shows the integrated profile over the entire system, obtained by integrating only the flux for which a velocity could be calculated, for each pixel, after smoothing the data to match the seeing, to minimise the noise in the profile. The comparison between the H$\alpha$ and HI curves extracted from \cite{2018ApJ...861...49H}
shows a good agreement between the distribution of warm and cold atomic hydrogen. It is noted that the HI distribution extends further in velocity than the H$\alpha$ distribution, which seems to imply that the southern region in HI is spatially more extended than the H$\alpha$ distribution, probably even further south, as the higher velocities seem to suggest.  
\cite{2007ggnu.conf..307F} described NGC 4098 as a system of two interacting spiral galaxies, which is itself interacting with the pair of spiral galaxies VV062a and VV062b. They observed the HI distribution in the USGC U451 group containing NGC 4098, revealing a very extended HI distribution that includes VV062 (see their Fig. 6). They placed lower limits on the HI content of NGC 4098, with S$_{HI}>$ 4.4 Jy km s$^{-1}$, providing an HI mass, 
$M_{HI}>9.9\times 10^{9}$ M$_\odot$.

\end{appendix}

\end{document}